\documentclass{aa}
\usepackage[varg]{txfonts}
\usepackage{amsmath,mathtools,multirow}
\usepackage{verbatim}
\usepackage{ulem} % for striking through text

\begin{document}

   \title{A low accretion efficiency of planetesimals\\ formed at planetary gap edges}

   \author{Linn E.J. Eriksson
          \inst{1}
          ,
          Thomas Ronnet \inst{1}
          , 
          Anders Johansen\inst{1,2}
          , 
          Ravit Helled\inst{3}
          , 
          Claudio Valletta\inst{3}
          , 
          Antoine C. Petit\inst{2}
          }

   \authorrunning{L.E.J. Eriksson}
   \institute{Lund Observatory, Department of Astronomy and Theoretical Physics, Lund University, Box 43, 221 00 Lund, Sweden \and Center for Star and Planet Formation, GLOBE Institute, University of Copenhagen, \O ster Voldgade 5-7, 1350 Copenhagen, Denmark \and Institute for Computational Science, University of Zurich, Winterthurerstr. 190, CH-8057 Zurich, Switzerland\\
              \email{linn@astro.lu.se}}

   \date{Received X; accepted X}

\abstract{Observations and models of giant planets indicate that such objects are enriched in heavy elements compared to solar abundances. The prevailing view is that giant planets accreted multiple Earth masses of heavy elements after the end of core formation. Such late solid enrichment is commonly explained by the accretion of planetesimals. Planetesimals are expected to form at the edges of planetary gaps, and here we address the question of whether these planetesimals can be accreted in large enough amounts to explain the inferred high heavy element contents of giant planets. We perform a series of N-body simulations of the dynamics of planetesimals and planets during the planetary growth phase, taking into account gas drag as well as the enhanced collision cross-section caused by the extended envelopes. We consider the growth of Jupiter and Saturn via gas accretion after reaching the pebble isolation mass and we include their migration in an evolving disk. We find that the accretion efficiency of planetesimals formed at planetary gap edges is very low: less than 10\% of the formed planetesimals are accreted even in the most favorable cases, which in our model corresponds to a few Earth-masses.  When planetesimals are assumed to form beyond the feeding zone of the planets, extending to a few Hill radii from a planet, accretion becomes negligible. Furthermore, we find that the accretion efficiency increases when the planetary migration distance is increased and that the efficiency does not increase when the planetesimal radii are decreased. Based on these results we conclude that it is difficult to explain the large heavy element content of giant planets with planetesimal accretion during the gas accretion phase. Alternative processes most likely are required, e.g. accretion of vapor deposited by drifting pebbles.} 

\maketitle 

\section{Introduction}
Matching interior structure models of Jupiter and Saturn to their measured gravity fields requires that the planets have a minimum heavy-element content of ${\sim} 20\, \textrm{M}_{\oplus}$, with upper bounds that are much higher and depend heavily on what model assumptions are used (e.g. \citealt{HelledGuillot2013,Wahl2017,Helled2018}).
Similarly, structure models of transiting planets find that extrasolar giants are typically enhanced in heavy elements, with estimated heavy element masses ranging from ${\sim} 10-100\, \textrm{M}_{\oplus}$ (\citealt{Guillot2006,MillerFortney2011,Thorngren2016}). Such large masses suggests that there are significant amounts of heavy elements in the H/He envelopes, indicating that giant planets typically have enriched atmospheres. Provided that envelope enrichment does not occur via erosion of the initial core alone \citep{Stevenson1982}, this implies that multiple Earth masses of heavy elements must have been accreted after the end of core formation.

Late heavy-element enrichment is often explained by the accretion of planetesimals (e.g. \citealt{Alibert2018}). In order to match the estimated heavy-element contents of giant planets, studies of this process typically have to assume a massive wide-stretched disk of planetesimals (e.g. \citealt{VenturiniHelled2020,Shibata2020}). However, simulations of planetesimal formation via the streaming instability (SI), which is one of the favored mechanisms for forming planetesimals \citep{Nesvorny2019}, suggests that planetesimals form in regions with locally enhanced solid-to-gas ratios (\citealt{YoudinGoodman2005,Johansen2009,Lyra2009,BaiStone2010,Carrera2015,Yang2017}). One such naturally occurring region is in the pressure bump generated at the edge of a planetary gap, where inwards drifting pebbles are trapped (\citealt{Stammler2019,Eriksson2020,Carrera2021}). 

In \citet{Eriksson2021} we studied the dynamical evolution of planetesimals formed at planetary gap edges, keeping the planet masses and locations fixed. We found that planetesimals are strongly scattered and leave their birthplace shortly after formation. In this study, we determine the amount of planetesimals that eventually collide with a planet, considering both planet migration and growth, and investigate whether the delivered mass is sufficiently high to explain the large heavy-element masses in giant planets. We therefore perform a suite of \textit{N}-body simulations, including the effect of gas drag on the planetesimals and the enhanced collision cross-section caused by the extended planetary envelope. We focus on  the formation of the Solar System's two gas-giant planets, Jupiter and Saturn, as both are massive enough to  open up deep gaps in the disk, and thus likely had planetesimals forming at their gap edges. We consider two different formation pathways for the planets, where one leads to a large-scale planetary migration, while the second one leads to a migration over a few au only. We continuously form planetesimals at the gap edges from the moment gas accretion initiates, i.e. when the pebble isolation mass has been reached ($M_{\rm iso}$, \citealt{Lambrechts2014,Bitsch2018}), until the time of disk dissipation. By further varying the planetesimal size and the formation location of the planetesimals relative to the planet, we explore  the sensitivity of the accretion efficiency to these parameters.

%We find that the maximum accretion efficiency in any simulation and onto any planet is $<10\%$. Given our assumed growth-tracks and using standard values for the disk model, the maximum amount of heavy elements delivered to Jupiter and Saturn is $3.1\, \textrm{M}_{\oplus}$ and $2.2\, \textrm{M}_{\oplus}$, respectively. These values are obtained by assuming that the entire flux of pebbles is converted into planetesimals at the gap edges, and thus represents an upper limit on the accreted planetesimal mass. The obtained accretion efficiencies does not vary with the planetesimal size; is higher in the case with a long attached phase and long migration distance; and are heavily dependent on the formation location of the planetesimals relative to the planet, such that there are close to zero collisions for planetesimals formed beyond the feeding zone of the planet. Based on these results we conclude that it is hard to explain the high heavy element content of giant planets with planetesimal accretion during the gas accretion phase. 

In Sect. \ref{sect: model}, we present our models for disk evolution and planet formation, as well as the setup of our simulations. We present the results of our simulations in Sect. \ref{sect: result}, where we show that the accretion efficiency of planetesimals formed at planetary gap edges is low, leading to the accretion of only a few Earth-masses of planetesimals in the most favorable cases. In Sect. \ref{sect: discussion section} we compare our results to the estimated heavy-element content of giant planets, and reach the conclusion that alternative processes most likely are required in order to explain the high heavy-element masses inferred from observations. The key findings of the paper are summarized in Sect. \ref{setc: conclusion}. Further information about our model and some additional figures can be found in Appendix \ref{sec: appendix disk model}-\ref{sec: appendix planetesmal accretion}.

\section{Model}\label{sect: model}
In this section we describe the ingredients and parameters of our simulations. We start by introducing our model for the structure and evolution of the protoplanetary disk. Planetesimals that move through the disk experience a drag force, which we account for following the method outlined in \citet{Eriksson2021}. Our model for planetary evolution contains planet migration, gas accretion and gap-opening. Collisions between planets and planetesimals are detected using a direct search method, which requires that the capture radius of the planet is known. We use the approximation from \citet{VallettaHelled2021} for planets that have attained a gaseous envelope (planets with masses above the pebble isolation mass), and assume that the capture radius is equal to the core radius for lower planetary masses. Finally we describe the numerical setup and initialization of our N-body simulations.

\subsection{Disk model}\label{subsec: disk model}
The evolution of the unperturbed surface density $\Sigma_{\rm unp}$ is modeled using a standard alpha-disk model, which is dependent on the disk accretion rate $\dot{M}_g$ and the scaling radius $r_{\rm out}$ \citep{LyndenBellPringle1974}. The kinematic viscosity is approximated as  
\begin{equation}
\nu = \alpha \Omega H^2,
\end{equation}
where $\alpha$ is the viscosity parameter, $\Omega$ is the Keplerian angular velocity, and $H$ is the scale height of the gaseous disk \citep{ShakuraSunyaev1973}. We take the scale height to be $H = c_{\textrm{s}}/\Omega$, where $c_{\textrm{s}}$ is the sound speed. The disk's midplane temperature was approximated using a fixed power-law structure for a passively irradiated disk \citep{ChiangGoldreich1997}. The disk accretion rate is given by the standard viscous accretion rate minus the rate at which gas is removed by photoevaporation $\dot{M}_{\rm pe}$, which we take to be a constant in time (see Appendix \ref{sec: appendix disk model} for details of our disk model).

Massive planets perturb their birth-disks by pushing material away from the vicinity of their orbits, leading to gap opening. We used a simple approach with Gaussian gap profiles to model these gaps. The height of the Gaussian is determined by the depth of the gap (see Section \ref{subsec: gap-profile}), and the width is taken to be one gas scale height at the planet location. Finally, in order to obtain the gas density at some height, $z,$ over the midplane we assume vertical hydrostatic equilibrium for the gas in the disk.

\subsection{Migration}\label{subsec:migration}
For the migration of planets we follow \citet{Kanagawa2018}, who found that embedded planets experience a torque which is equal to the classical type-I torque multiplied by the relative gap height, resulting in the migration rate:
\begin{equation}
\dot{r} = \dot{r}_{\rm I} \times \frac{\Sigma_{\rm{gap}}}{\Sigma_{\rm unp}}
\end{equation}
(see Eq. 3-4 in \citet{Johansen2019} for the classical type-I migration rate).
This results in little or no migration before the planet reaches a few Earth masses, fast migration during the last stages of pebble accretion and the first stages of gas accretion, and slow migration during runaway gas accretion when the gap has become deep. For the implementation of migration into the N-body code we follow \citet{CresswellNelson2008}, who provides the timescales for radial migration as well as the associated eccentricity and inclination damping. The corresponding accelerations are then applied directly to the planets at fixed time-intervals.

\subsection{Gas accretion}\label{subsec: gas accretion}
Protoplanets can accrete gaseous envelopes already during the early phases of core-formation.  During this stage the heating of the envelope from the flux of pebbles prevents it from contracting onto the planet. At a later stage, when the protoplanet reaches the pebble isolation mass and the flux of pebbles stops, the envelope contracts. This is the first stage of significant gas accretion; however, already before this stage some small amounts of highly polluted gas can become bound to the protoplanet inside its Hill sphere \citep{Bitsch2015}. We follow the assumption of \citet{Bitsch2015} that 10\% of the material accreted by the planet prior to the pebble isolation mass is in gas, so that the final mass of the core is $0.9\times M_{\rm iso}$. Note that this is a simplification, and in reality the gas fraction is expected to increase as the core grows more massive \citep{VallettaHelled2020}.  

The gas accretion rate of a protoplanet during the contraction phase (also known as the attached phase) is highly uncertain and there are several  prescriptions to describe this stage, many of which are vastly different. In this work we use two different prescriptions:  scheme 1 results in a short contraction phase while scheme 2 leads to a relatively long contraction timescale.  
In scheme 1 we follow the gas accretion model outlined in \citet{Johansen2019} (Eq. 38-40). The envelope's contraction occurs on the Kelvin-Helmholtz timescale, which depends on the planetary mass $M_p$ and the envelope's opacity $\kappa$ \citep{Ikoma2000}. For large planetary masses, this accretion rate becomes higher than what the disk can supply. At this point the planet enters the runaway gas accretion phase (also known as the detached phase), and the growth becomes limited by the rate at which gas can enter the Hill sphere \citep{TanigawaTanaka2016}. As an additional constraint, the planet can not accrete at a rate which is higher than the global disk accretion rate $\dot{M}_g$. Furthermore, \citet{LubowD'Angelo2006} finds that even high mass planets can not block all gas from passing the gaps, and that the protoplanets do not accrete at a rate higher than 75-90\% of the disk's accretion rate. Therefore, we limit the maximum gas accretion rate to 80\% of the disk's accretion rate. The gas accretion rate onto the planet in scheme 1 is then taken to be the minimum of the three aforementioned rates. 

In scheme 2 we roughly follow the gas accretion model which is used in \citet{Bitsch2015}. During the contraction phase, here defined as $M_{\rm env}<M_{\rm core}$, the envelope contracts on a long time-scale while accreting some gas. The accretion rate during this phase depends on the envelope opacity, the core density $\rho_{\rm core}$, the mass of the core and envelope, and the disk temperature (see Eq. 17 in \citet{Bitsch2015}, originally derived in \citet{PisoYoudin2014}). Runaway gas accretion initiates when $M_{\rm env}>M_{\rm core}$, and for this phase we use the gas accretion rate from \citet{TanigawaTanaka2016}, which is the same as in scheme 1. Furthermore, we also limit gas accretion to 80\% of the disk accretion rate throughout the entire process. The dependence of all aforementioned gas accretion rates on the planetary mass is plotted in Fig. \ref{fig: gasAccretion}. 

\subsection{Gap-depth}\label{subsec: gap-profile}
For the relative depth of the planetary gaps we follow \citet{Johansen2019}, who showed that the gap-depth scales with the pebble isolation mass as: 
\begin{equation}
\frac{\Sigma_{\rm gap}}{\Sigma_{\rm unp}} = \frac{1}{1+\left(\frac{M_p}{2.3M_{\rm iso}}\right)^2}.
\end{equation}
We use the fit from \citet{Bitsch2018} to calculate the pebble isolation mass, which depends on the turbulent viscosity $\alpha_t$ and the unperturbed radial pressure gradient of the disk $\partial \ln P/\partial \ln r$. We use an unperturbed surface density gradient of $-15/14$, which results in a radial pressure gradient of $-2.7857$.

\begin{comment}
\begin{equation}
\begin{split}
M_{\rm iso} = & 25M_{\oplus}\left[\frac{H/r}{0.05}\right]^3 \left[0.34 \left(\frac{\log(\alpha_3)}{\log(\alpha_t)}\right)^4 + 0.66\right] \\
& \times \left[1-\frac{\partial \ln P/\partial \ln r + 2.5}{6}\right],
\end{split}
\end{equation}
where $\alpha_3=10^{-3}$, $\alpha_t$ is the turbulent viscosity and $\partial \ln P/\partial \ln r$ is the unperturbed radial pressure gradient of the disk. 
\end{comment}

\subsection{Numerical setup and initialization}\label{subsec: nomerical setup and initialization}
\begin{table}
\centering
\caption{Parameters used throughout the simulations. The parameters which are varied are marked with $*$.}
\label{table:parameters}
\begin{tabular}{lll}
\hline\hline
Parameter             & Value                                               &                                    \\ \hline
\multicolumn{3}{c}{Planet}                                                                                       \\ \hline
$e_0$                 & $10^{-3}$                                           & initial eccentricity               \\
$i_0$                 & $10^{-3}$                                           & initial inclination                \\
$\rho_{\rm core}$     & $5.5 \times 10^{3}\, \textrm{kg}\, \textrm{m}^{-3}$ & density of core                    \\
$\kappa$              & $5\times 10^{-2}\, \textrm{m}^2\, \textrm{kg}^{-1}$ & opacity of envelope                \\ \hline
\multicolumn{3}{c}{Planetesimal}                                                                                 \\ \hline
$N_{\rm pl}$          & $10^4$                                              & number of planetesimals            \\
$R_{\rm pl}$          & $10^5, 10^4, 300\, \textrm{m}$           & radius$^*$                             \\
$\rho_{\rm pl}$       & $10^{3}\, \textrm{kg}\, \textrm{m}^{-3}$   & density                            \\
$a_{\rm pl}$          & $2-3, 4-5\, R_H$                                  & initial location rel planet$^*$ \\
$e_{\rm pl,0}$          & $\sim 5\times 10^{-3}$                                  & initial eccentricity \\
$i_{\rm pl,0}$          & $\sim 5\times 10^{-3}$                                  & initial inclination \\ \hline
\multicolumn{3}{c}{Disk}                                                                                         \\ \hline
$r_{\rm out}$         & $75\, \textrm{au}$                                  & scaling radius                    \\
$\dot{M}_g(t=0)$      & $10^{-7}\, M_{\odot}\, \textrm{yr}^{-1}$            & initial disk accretion rate        \\
$\dot{M}_{\rm photo}$ & $5.9\times 10^{-9}\, M_{\odot}\, \textrm{yr}^{-1}$  & rate of photo evaporation          \\
$t_{\rm evap}$        & $3\, \textrm{Myr}$                                  & disk lifetime                      \\
$\alpha$              & $10^{-2}$                                           & viscosity parameter                \\
$\alpha_{\rm t}$      & $10^{-4}$                                           & turbulent parameter      \\
$\epsilon$            & 1\%                                                 &solid-to-gas ratio                                    \\ \hline
\multicolumn{3}{c}{Grid}                                                                                         \\ \hline
$N_{\rm grid}$        & 1000                                                & number of gridcells                \\
$r_1$                 & $0.1\, \textrm{au}$                                 & innermost grid point               \\
$r_{1000}$            & $100\, \textrm{au}$                                 & outermost grid point               \\ \hline
\multicolumn{3}{c}{Simulation}                                                                                   \\ \hline
dt                  & $0.6\, \textrm{yr}$                                 & WHFAST timestep                    \\
$t_{\rm max}$         & $10\, \textrm{Myr}$                                 & simulation time                    \\ \hline
\end{tabular}
\end{table}

Table \ref{table:parameters} lists the parameter values used in our simulations.
The simulations were performed with the $N$-body code REBOUND and executed using the hybrid symplectic integrator MERCURIUS (\citealt{ReinLiu2012,Rein2019}). The WHFAST timestep was set to be one twentieth of Jupiter's current
dynamical timescale, and we only performed disk evolution, migration and gas accretion on this timestep. The planets and the central star were added as active particles, and the planetesimals as test particles. We used a central star of solar mass and solar luminosity. Collisions were detected using a direct search algorithm and resulted in perfect merging. Particles that leave the simulation domain, which is centered on the sun and stretches $100\, \textrm{au}$ in $x$ and $y$-direction and $20\, \textrm{au}$ in $z$-direction, are recorded and removed from the simulation. 

The protoplanetary disk was modeled using a linear grid with 1000 grid cells, stretching from $0.1$ to $100\, \textrm{au}$. We used standard values of $75\, \textrm{au}$ for the scaling radius and $10^{-2}$ for the viscosity parameter. The initial disk accretion rate was set to $10^{-7}\, M_{\odot}\, \textrm{yr}^{-1}$, and we chose the photoevaporation rate such that the disk obtained a lifetime of $3\, \textrm{Myr}$. We continued the orbital integration for an additional $7\, \textrm{Myr}$ after disk dispersal. 

We modeled Jupiter's formation starting at the pebble isolation mass, and for Saturn we also included a prescription for the growth of the core. The reason is that in our model Jupiter reaches the pebble isolation mass and begins to form planetesimals at the gap edge earlier than Saturn, and while we assume that there is no planetesimal formation at Saturn's gap edge during core formation, the formation of its core could still dynamically effect the planetesimals formed at Jupiter's gap edge. The initial semimajor axis and formation time of the planets are found in Section \ref{subsec: growthTrack}. The initial orbital eccentricity and inclination were set to $10^{-3}$, and we used a constant opacity of $5\times 10^{-2}\, \textrm{m}^2\, \textrm{kg}^{-1}$ for the gaseous envelope.

Each simulation contained a total of 10,000 planetesimals, which were injected at the gap edges ten at the time from the beginning of the simulation until disk dissipation. Given the formation times stated in Table \ref{table:growth-track}, this means that we injected 10 planetesimals into the simulation roughly every 1,000 year.  The planetesimal radii was varied in between the simulations and set to either $300\, \textrm{m}$, $10\, \textrm{km}$ or $100\, \textrm{km}$ (\citealt{Bottke2005,Morbidelli2009,Johansen2015}). The exact formation location of planetesimals at gap edges is unknown \citep{Carrera2021}. Here we tried two different formation locations: in the first setup we initiated the planetesimals uniformly between 2 and 3 Hill radii from the planets (thus interior of the single planet feeding zone at $2\sqrt{3}\, R_H$); and in the second between 4 and 5 Hill radii from the planets (thus exterior of the feeding zone). The feeding zone is defined as the region where the Jacobi energy of the planetesimals is positive \citep{ShiraishiIda2008}. Finally, in order to provide better statistics we performed three simulations per parameter set.

\section{Result}\label{sect: result}

\subsection{Planet growth-tracks}\label{subsec: growthTrack}

\begin{figure*}
\centering
{\includegraphics[width=\columnwidth]{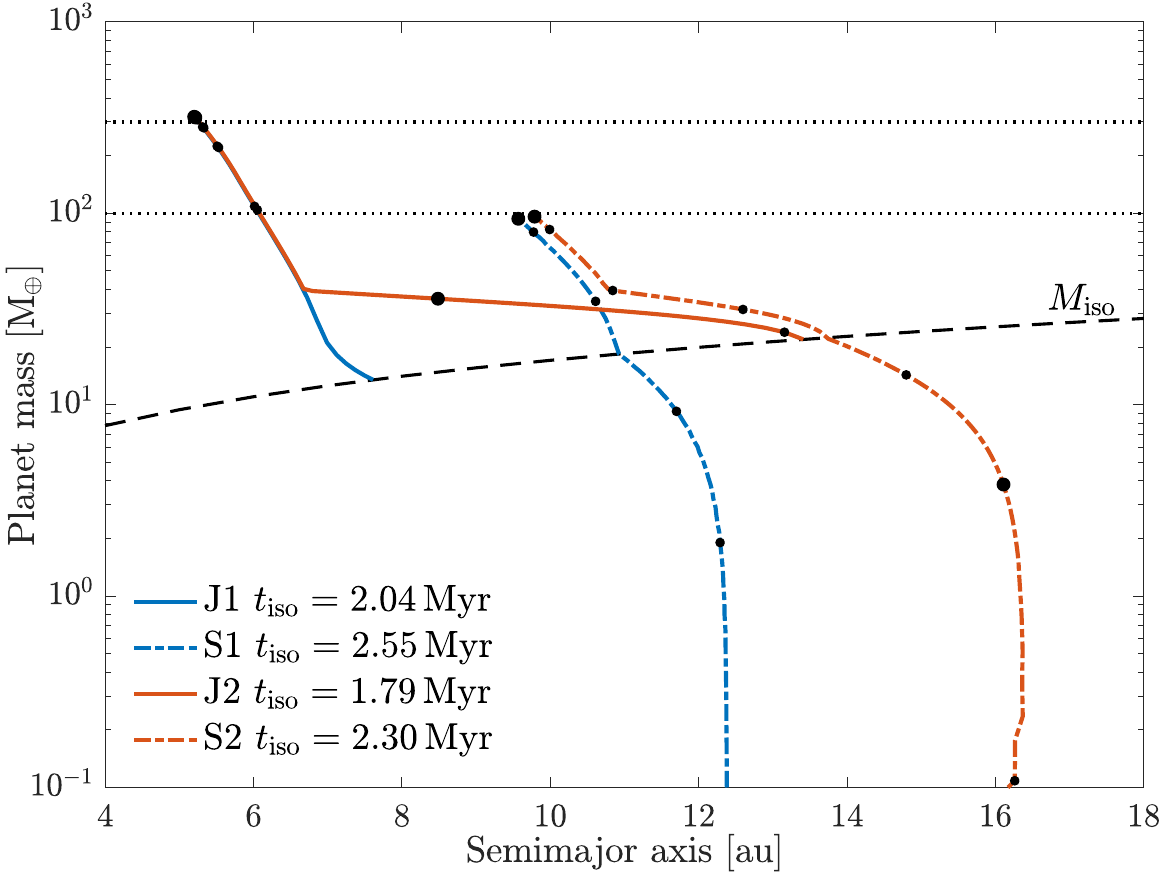}}
{\includegraphics[width=\columnwidth]{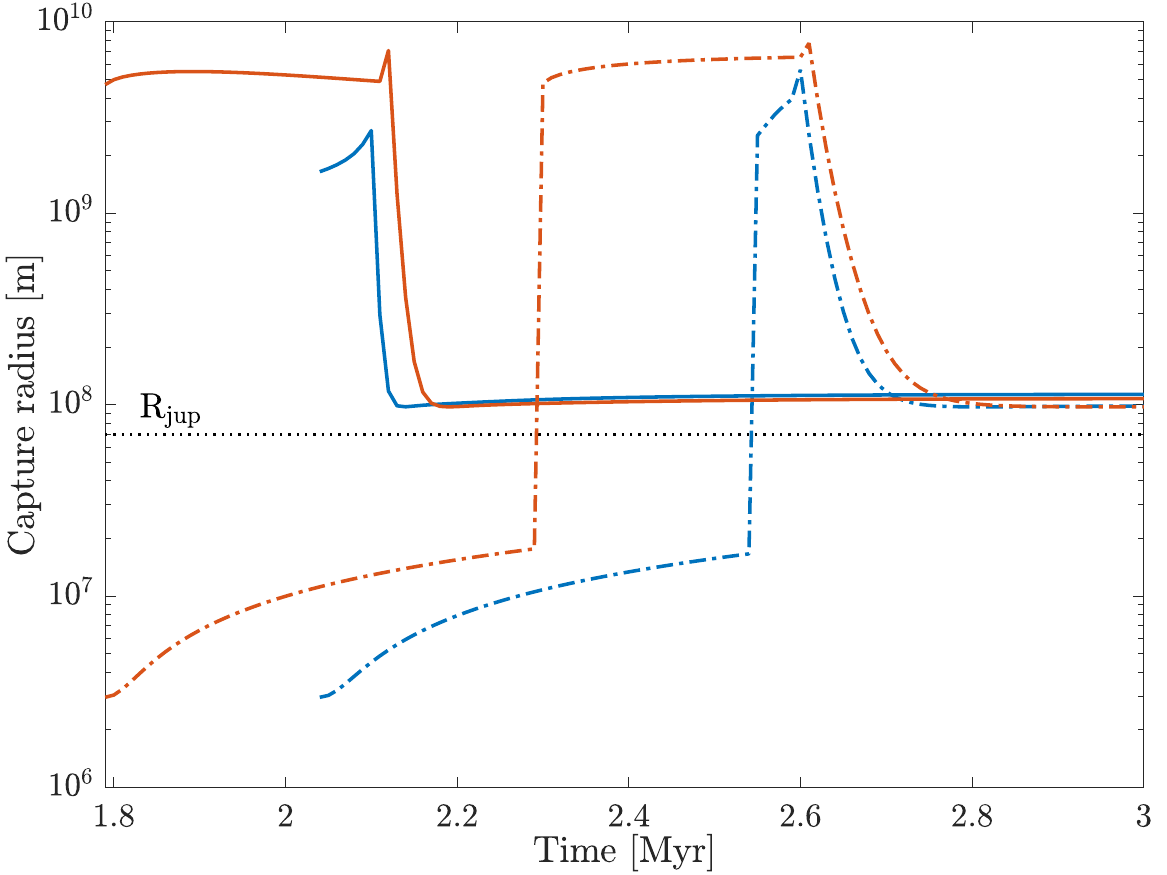}}
\caption{\textit{Left}: Planet mass versus semimajor axis evolution for Jupiter (J) and Saturn (S) in gas accretion scheme 1 and 2. Large dots indicate a time of 2 and $3\, \textrm{Myr}$, and small dots are separated by $0.2\, \textrm{Myr}$. We start the simulation at the time when Jupiter reaches the pebble isolation mass ($t_{\rm iso}$), and further include the growth and migration of Saturn's core. In scheme 1 the contraction of the envelope occurs on a short timescale, resulting in fast gap-opening and little migration. In scheme 2 this phase is significantly longer, and thus the planets migrate a further distance before coming to a halt due to gap-opening. \textit{Right}: Capture radius versus time for the growth-tracks presented in the left panel, calculated using a planetesimal radius of $100\, \textrm{km}$. The approximation for the capture radius has 3 regimes: before gas accretion initiates the capture radius is equal to the core radius; during the first phase of gas accretion the envelope is enhanced, resulting in a large capture radius; after the onset of runaway gas accretion the capture radius decreases by roughly two orders of magnitude, and takes on a value of about 1.5 times the current Jupiter radius.}
\label{fig: growthTrack}
\end{figure*}

\begin{table}
\centering
\caption{Parameters for the planet growth-tracks. $t_{\rm iso}$ and $a_{\rm iso}$ are the time and semimajor axis at which the planet reaches the pebble isolation mass $M_{\rm iso}$. $a_{\rm core}$ is the semimajor axis at which Saturn's core has to be initiated in order for it to reach $a_{\rm iso}$ at time $t_{\rm iso}$.}
\label{table:growth-track}
\begin{tabular}{lllll}
\hline\hline
Run & $t_{\rm iso}$         & $a_{\rm iso}$        & $M_{\rm iso}$                 & $a_{\rm core}$        \\ \hline
J1  & $2.04\, \textrm{Myr}$ & $7.61\, \textrm{au}$ & $13.50\, \textrm{M}_{\oplus}$ &                       \\
S1  & 2.55                  & 10.94                & 18.42                         & $12.38\, \textrm{au}$ \\
J2  & 1.79                  & 13.41                & 21.93                         &                       \\
S2  & 2.30                  & 13.52                & 22.09                         & 16.17                 \\ \hline
\end{tabular}
\end{table}

Given our models for migration and gas accretion (introduced in Section \ref{subsec:migration} and \ref{subsec: gas accretion}) we searched for growth-tracks that resulted in Jupiter and Saturn having their current mass and semimajor axis at the time of disk dissipation. The resulting time and semimajor axis at which the planets begin to migrate and accrete gas, as well as the corresponding pebble isolation mass, can be found in Table \ref{table:growth-track}. According to these results, Saturn reaches the pebble isolation mass much later than Jupiter. We begin our simulations at the time when Jupiter reaches the pebble isolation mass, and although we assume that planetesimals do not form at Saturn's gap edge during core formation, the formation of Saturn's core could still dynamically affect the planetesimals formed at Jupiter's gap edge. Therefore we also include the growth and migration of Saturn's core in our simulations. 

Rather than calculating the actual pebble accretion rate at each iteration, we used a simple approach for Saturn's core growth. We initiated the core at $0.1\, \textrm{M}_{\oplus}$ at the beginning of the simulation, and grew it up to $M_{\rm iso}$ following a reasonable growth rate, which we picked to be $M_{\rm core} \propto t^2$. This growth rate is slightly shallower than the growth rate suggested by pebble accretion ($M_{\rm core} \propto t^3$ in the 3D regime, see e.g. \citealt{Morbidelli2015}); however, the effect on the simulation outcome should be negligible (as will be shown in Section \ref{sect: result}, accretion onto Jupiter happens before the planetesimals have any chance to dynamically interact with Saturn's core).  Note that during this phase, 10\% of the accretion is assumed to be in the form of gas, so that the actual core mass is 90\% of the above mass (see Section \ref{subsec: gas accretion}). Finally we also consider the migration of the core, which is performed using our normal migration prescription. The semimajor axis where the core needs to be initiated in order for Saturn to reach the pebble isolation mass at the right location is listed in the last column of Table \ref{table:growth-track}.

The resulting growth-tracks are presented in the left panel of Fig. \ref{fig: growthTrack}, where 1 and 2 denote gas accretion scheme 1 and gas accretion scheme 2, respectively. In scheme 1 the envelope's contraction occurs on a relatively short time-scale, leading to quick gap-opening and little migration. In scheme 2 the contraction phase is significantly longer, and as a result the planet migrates much farther before runaway gas accretion initiates. Gravitational perturbations from Jupiter cause some variations in the growth-track of Saturn's core, but these are too small to have any impact on the simulation outcome. The corresponding time-evolution of the capture radius for these growth-tracks is presented in the right panel of Fig. \ref{fig: growthTrack}. The capture radius during gas accretion is calculated using the approximation from \citet{VallettaHelled2021}, which has two regimes depending on if the planet is in the attached phase ($M_{\rm env}<M_{\rm core}$) or the detached phase ($M_{\rm env}>M_{\rm core}$). For the detached phase we use the fit obtained at $10^7\, \textrm{yr}$. During the attached phase the envelope is enhanced, resulting in a large capture radius. This phase ends once runaway gas accretion initiates, and the capture radius decreases to about 1.5 times the current Jupiter radius. 

\subsection{Dynamical evolution of planetesimals}

\begin{figure*}
\centering
{\includegraphics[width=2\columnwidth]{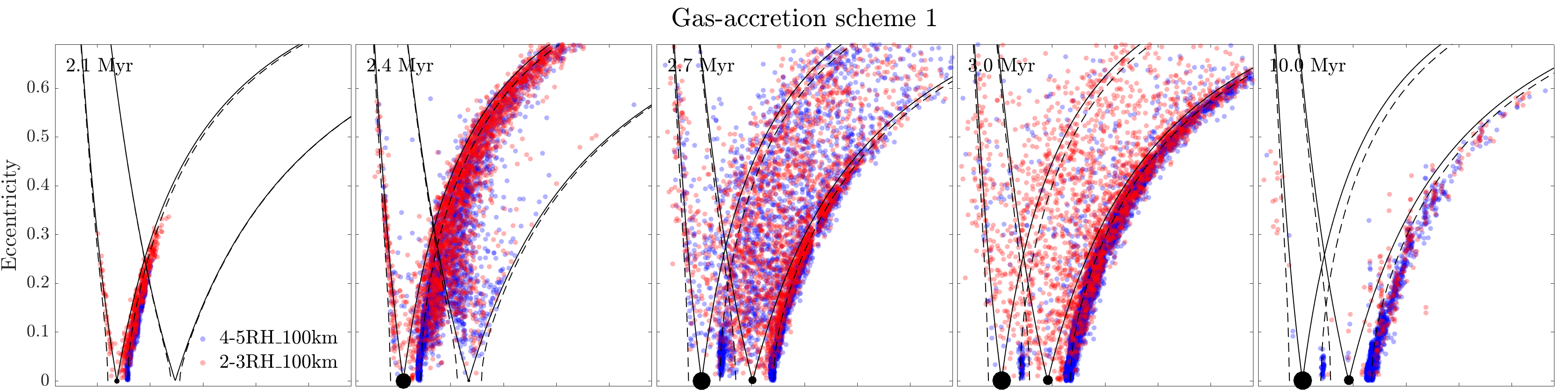}}
{\includegraphics[width=2\columnwidth]{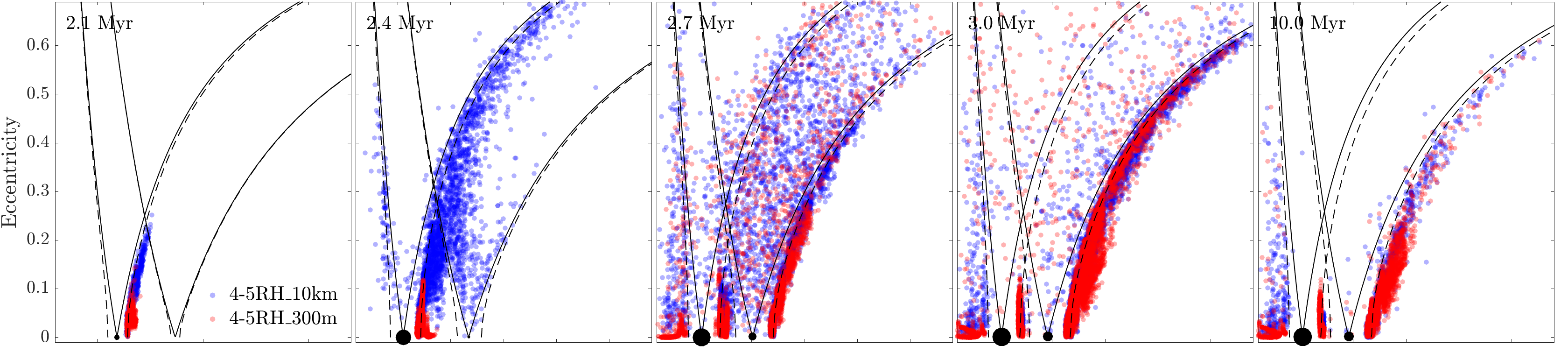}}
{\includegraphics[width=2\columnwidth]{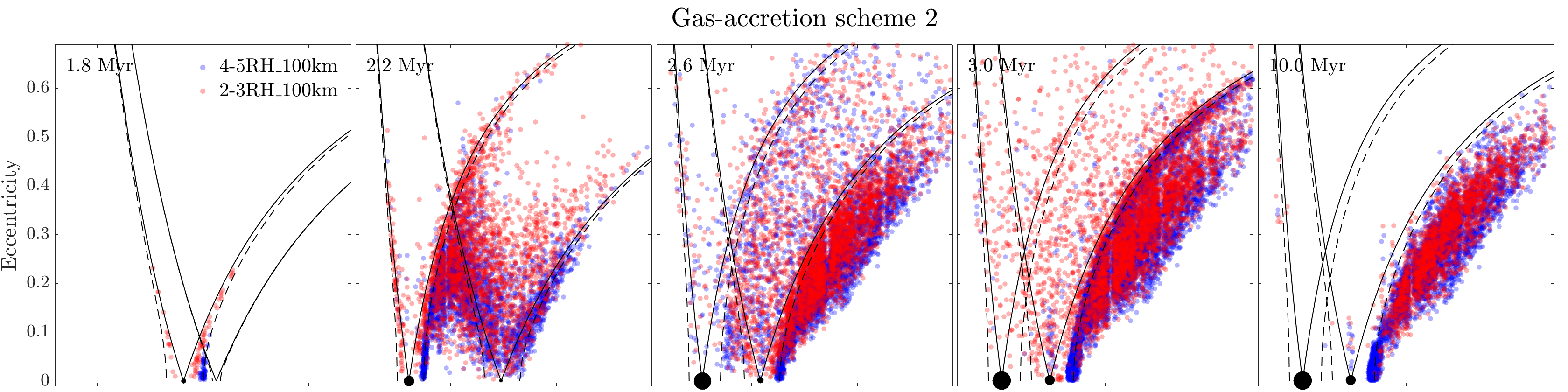}}
{\includegraphics[width=2\columnwidth]{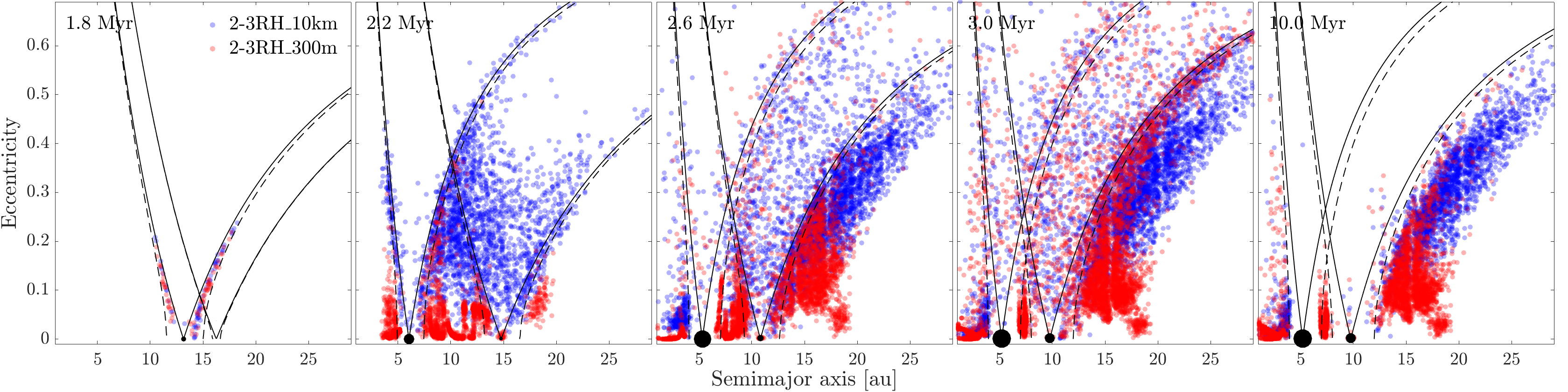}}
\caption{Eccentricity and semimajor axis evolution for the planetesimals in some selected simulations. Each simulation has a total of 10,000 planetesimals being injected continuously until the time of disk dissipation. The black dots indicate the current mass and location of Jupiter and Saturn. The black lines are lines of equal Tisserand parameter going through the planet location (solid line) and the planet location offset by 5 Hill radii (dashed line). Planetesimals initiated inside the feeding zone of the planet (labeled $2-3\, R_{\rm H}$) suffer stronger and faster scattering than those initiated outside the feeding zone (labeled $4-5\, R_{\rm H}$). Decreasing the planetesimal size leads to more gas-drag and lower eccentricities, and results in a population of circularized planetesimals interior of Jupiter.  }
\label{fig: eVSa}
\end{figure*}

\begin{figure*}
\centering
{\includegraphics[width=1\columnwidth]{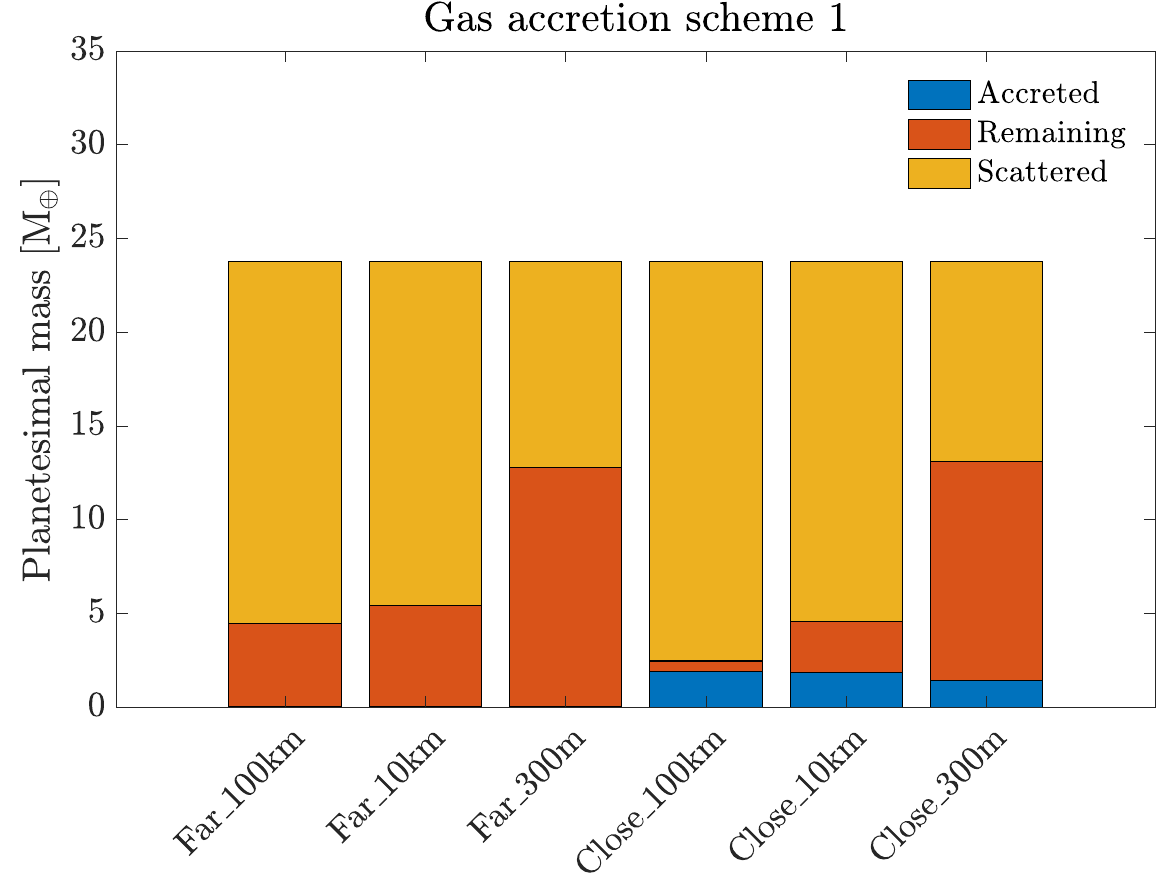}}
{\includegraphics[width=1\columnwidth]{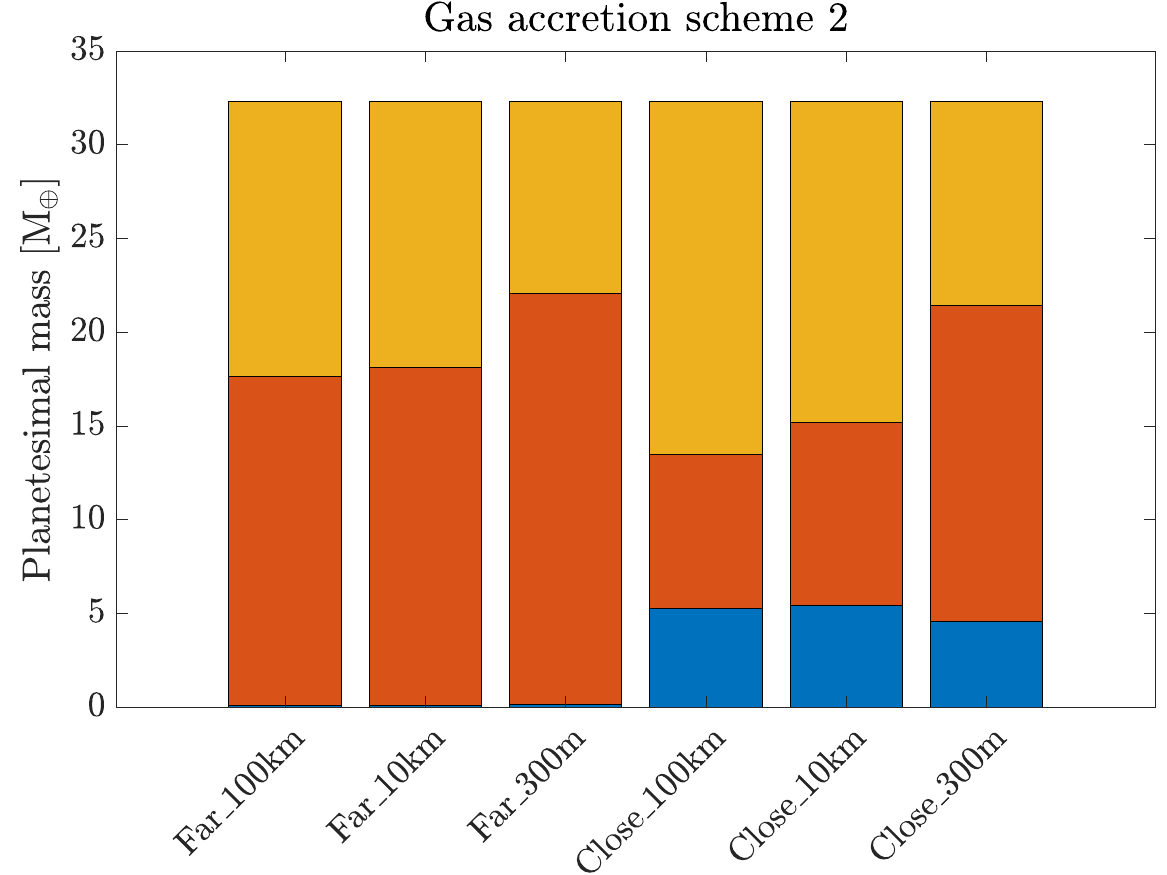}}
\caption{Histogram showing the total planetesimal mass that has been either accreted onto the planets, remains in the system at the end of the simulation, or been scattered beyond the simulation domain for all simulations in gas accretion scheme 1 (\textit{left}) and gas accretion scheme 2 (\textit{right}). Small planetesimals experience more efficient gas drag than large planetesimals, and are retained in the system at a higher rate. }
\label{fig: histMass}
\end{figure*}

Planetesimals are continuously injected into the simulation until the time of disk dissipation, following the procedure outlined in Section \ref{subsec: nomerical setup and initialization}. We make two important assumptions regarding planetesimal formation: (1) there is no planetesimal formation at the gap edges before the pebble isolation mass has been reached; (2) all pebbles that reach the gap edges are trapped and immediately converted into planetesimals. Taken together, this means that when Saturn reaches the pebble isolation mass the drift of pebbles towards Jupiter is terminated. The time it takes for the remaining pebbles to reach Jupiter's gap edge after this is relatively short, only $~10^3-10^4\, \textrm{yr}$, and therefore we assume that planetesimals cease to form at Jupiter's gap edge when Saturn reaches the pebble isolation mass. 
Fig. \ref{fig: eVSa} shows the time evolution of the eccentricities and semimajor axes of the planetesimal orbits, and how they vary with: 1) the initial formation location relative to the planet; 2) the planetesimal size; and 3) the different gas accretion schemes. Lines of equal Tisserand parameter for coplanar orbits are included in the Figure and can be used to understand the planet scattering.

The effect of varying the initial formation location of the planetesimals is shown in row 1 and 3 of Fig. \ref{fig: eVSa}. Initiating the planetesimals inside the feeding zone of the planet results in faster and stronger scattering compared to the case when they are initiated outside of the feeding zone. The number of planetesimals scattered into the inner Solar System is also significantly higher in the former case (see also Fig. \ref{fig: massInt4au}). The amount of planetesimals which are scattered beyond the simulation domain does however not vary with the formation location, which can be seen in Fig. \ref{fig: histMass}. Regardless of the formation location, most planetesimals that remain in the system past disk dispersal are located in a disk beyond Saturn.

In row 2 and 4 of Fig. \ref{fig: eVSa} we show how the dynamical evolution changes when we decrease the planetesimal size. Smaller planetesimals are more affected by gas drag than larger planetesimals, and therefore have their eccentricities more damped. Interior of Jupiter the gas density is high enough to circularize small planetesimals, resulting in a stable population of planetesimals that remain until the end of the simulation. This population has low eccentricities in the case of $300\, \textrm{m}$-sized planetesimals; eccentricities up to around 0.4 in the case of $10\, \textrm{km}$-sized planetesimals; and does not exist at all in the case of $100\, \textrm{km}$-sized planetesimals. Furthermore there is a small population of planetesimals located around the outer 3:2 resonance with Jupiter, which appears early on and remains until $10\, \textrm{Myr}$. As expected, the number of planetesimals that are left in the system at the end of the simulation increases with decreasing planetesimal size, which is because of the stronger gas drag (see Fig. \ref{fig: histMass}).

Finally, the effect of changing the gas accretion scheme can be seen by comparing the two upper rows in Fig. \ref{fig: eVSa} with the two lower rows. Since the planets migrate larger distances in scheme 2, planetesimals form in a wider region of the disk. A significant amount of the planetesimals formed at Saturn's gap edge become detached from the scattering region due to the planet migration (recognized by being located beyond and below the equal Tisserand parameter lines), resulting in a broader planetesimal disk in scheme 2. Towards the end of the simulation there are more planetesimals remaining in the system with gas accretion scheme 2 (see Fig. \ref{fig: histMass}).

\subsection{Planetesimal accretion}

\begin{figure*}
\centering
{\includegraphics[width=1\columnwidth]{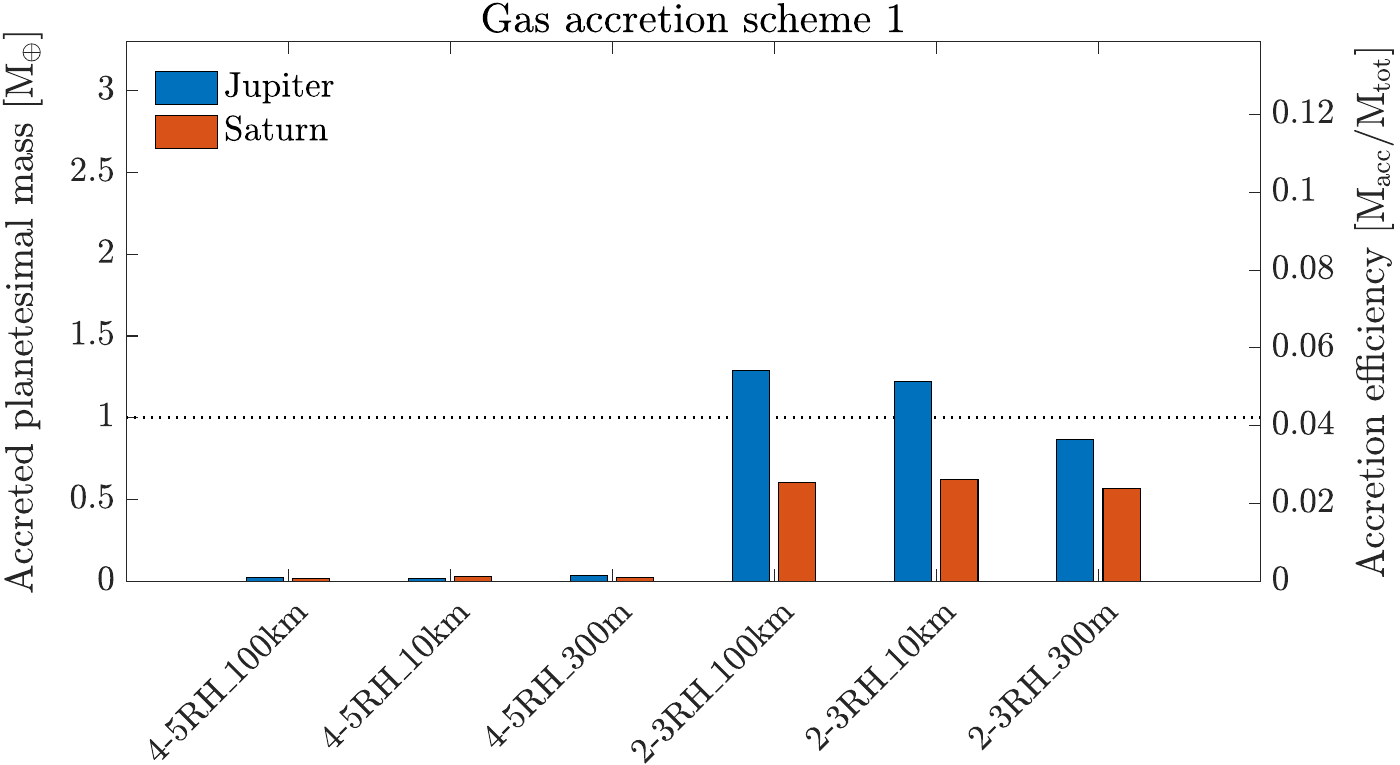}}
{\includegraphics[width=1\columnwidth]{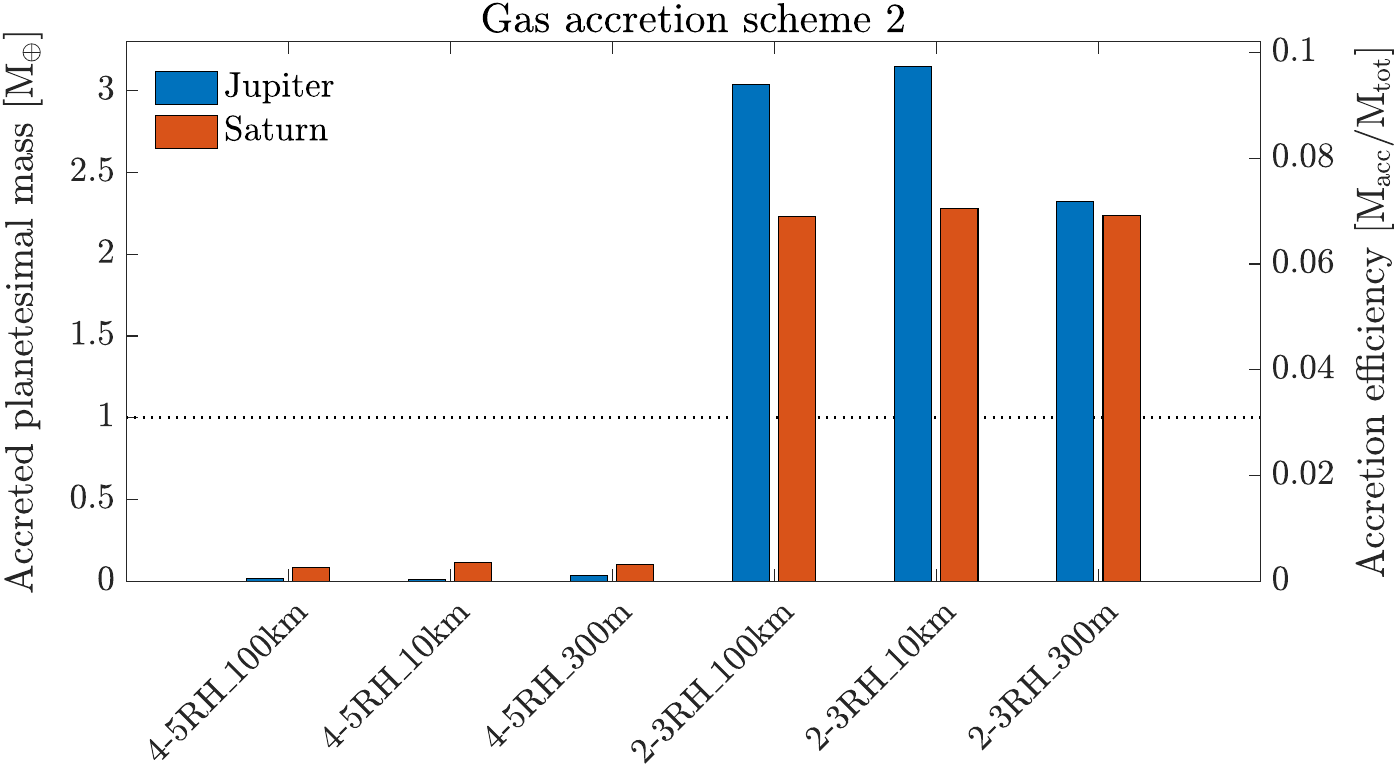}}
\caption{Total accreted planetesimal mass onto Jupiter and Saturn, and the corresponding accretion efficiency, for all simulations in gas accretion scheme 1 (\textit{left}) and gas accretion scheme 2 (\textit{right}). The accretion efficiency is much higher for planetesimals formed inside the feeding zone, and does not change significantly with planetesimal size. The maximum accretion efficiency obtained for any planet is $<10\%$, which corresponds to a mass of ${\sim} 3\, \textrm{M}_{\oplus}$. }
\label{fig: colDiag}
\end{figure*}

In order to calculate the mass represented by each super-particle, we assume that the pebbles follow the viscous evolution of the disk \citep{Johansen2019}, and that all pebbles which reach the gap edge are converted into planetesimals. The planetesimal formation rate is then simply 1\% of the disk accretion rate (note that we do not consider photo-evaporation in this calculation as it only affects the gas component of the disk), and the total planetesimal mass that forms in the simulation is calculated by integrating $0.01 \times \dot{M}_{\rm g}$ from $t_{\rm iso, jup}$ to $t_{\rm evap}$. With these assumptions we form $23.8\, \textrm{M}_{\oplus}$ of planetesimals in scheme 1 and $32.3\, \textrm{M}_{\oplus}$ in scheme 2. The total planetesimal mass formed at each individual gap edge is $14.1\, \textrm{M}_{\oplus}$ for Jupiter in scheme 1, $9.7\, \textrm{M}_{\oplus}$ for Saturn in scheme 1, $16.2\, \textrm{M}_{\oplus}$ for Jupiter in scheme 2 and $16.1\, \textrm{M}_{\oplus}$ for Saturn in scheme 2. Since the disk accretion rate decreases with time, the super-particles that form in the beginning of the simulation will be much more massive than those which form towards the end. Given the assumed solid-to-gas ratio, disk model and formation time of Jupiter the above masses represents an upper limit on the formed planetesimal mass. 

In Fig. \ref{fig: histMass} we show for each simulation how much of the total planetesimal mass that has been either accreted onto the planets, remains in the system or has been scattered beyond the simulation domain at the end of the simulation. Fig. \ref{fig: colDiag} shows how much of the planetesimal mass that has been accreted onto Jupiter and Saturn, along with the corresponding accretion efficiency. The first thing to be noticed is that the maximum accretion efficiency in any simulation and for any planet is $<10\%$, and the highest amount of solid material accreted onto Jupiter and Saturn is $3.2\, \textrm{M}_{\oplus}$ and $2.3\, \textrm{M}_{\oplus}$, respectively. This shows that planetesimal accretion during the gas accretion phase of giant planet formation is a very inefficient process.

Comparison of the simulation results shows that initializing the planetesimals inside the feeding zone of the planet results in a lot more collisions than placing them outside of the feeding zone. This is partly because a fraction of the planetesimals formed within the feeding zone are captured immediately after formation due to the enhanced envelope (see Fig. \ref{fig: growthTrack} and \ref{fig: collTime_formLoc}). The planetesimals formed beyond the feeding zone are too far away from the planet to be affected by the enhanced envelope, and suffer only strongly unfocused collisions later on during the evolution. 

As mentioned above, decreasing the planetesimal size results in more gas-drag. Furthermore, it also results in an increased capture radius of the planets during the enhanced envelope phase \citep{VallettaHelled2021}, as smaller planetesimals can be captured further up in the atmosphere. The combined effect on the accretion efficiency of planetesimals, however, turns out to be small. In the case of Jupiter, we find that $300\, \textrm{m}$-sized planetesimals are accreted at a ~30\% lower rate than 10 and $100\, \textrm{km}$-sized planetesimals when formed inside the feeding zone. This is because if the planetesimals are not accreted immediately, they are scattered and eccentricity damping through gas-drag quickly puts them out of the feeding zone, resulting in fewer planetesimals crossing Jupiter's orbit. In the case of Saturn we find no difference between using small and large planetesimals. Some authors have suggested that using small planetesimals might solve the problem of the low planetesimal accretion efficiency (e.g. \citealt{Alibert2018}), but our results show that this is not the case for planetesimal accretion during the gas accretion phase.

When comparing the accretion efficiencies in scheme 1 and 2 we find that it is generally higher in scheme 2. This is because the enhanced envelope phase lasts longer in scheme 2, allowing for more planetesimals to be captured just after formation (see Fig. \ref{fig: growthTrack}). Finally, in most of our simulations, Jupiter accreted a higher planetesimal mass than Saturn. In Fig. \ref{fig: planAcc} we show how the cumulative accreted planetesimal mass evolves with time.

\section{Discussion}\label{sect: discussion section}

\subsection{The fate of removed planetesimals}\label{subsec: removed planetesimals}

\begin{figure}
\centering
{\includegraphics[width=\columnwidth]{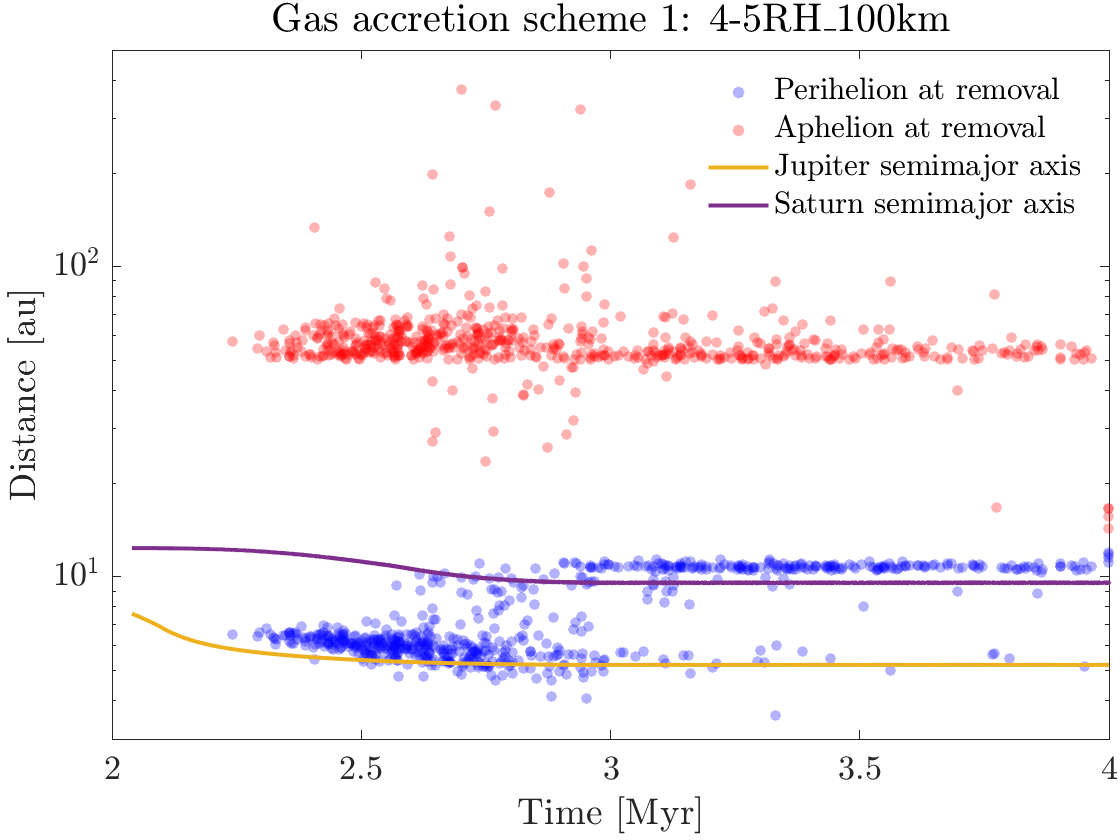}}
\caption{Plot showing the perihelion/aphelion of 1000 planetesimal orbits at the last time-step before they are scattered beyond the simulation domain and thus removed from the simulation, for one of the simulations in gas accretion scheme 1. The time evolution of the planet semimajor axes is included on the plot. The majority of the scattered planetesimals have an aphelion of ${\sim} 50\, \textrm{au}$ and a perihelion located ${\sim} 1\, \textrm{au}$ exterior of the planets, suggesting that they are not strongly scattered but rather are part of a scattered disc where eccentricities and aphelia increase gently with time.}
\label{fig: distVSt_removed}
\end{figure}

When a planetesimal is scattered beyond the simulation domain it is removed from the simulation. In order to get an idea of how the planetesimal would have evolved given a larger simulation domain, we studied how the planetesimal orbits looked like just before they are removed. The perihelion and aphelion for 1000 planetesimal orbits at the last time-step before removal is showed in Fig. \ref{fig: distVSt_removed}. Since most planetesimals have aphelion around $50\, \textrm{au}$ and perihelion located ${\sim} 1\, \textrm{au}$ exterior of the planets, this indicates that they still belong to the planetary scattered disks at the time of removal. In other words, they do not suffer strong planetary encounters and should not have been ejected from the system. It is therefore likely that these planetesimals would have remained in or around the scattered disk also at $10\, \textrm{Myr}$, had the simulation domain been larger. If this is the case, then the amount of planetesimal mass deposited in the region exterior of Saturn is at least $10-20\, \textrm{M}_{\oplus}$. 

\subsection{Implications for Solar System formation}\label{subsec: implications to SS formation}
The estimated total heavy-element mass in Jupiter and Saturn is $\gtrsim 20\, \textrm{M}_{\oplus}$, with upper bounds that are much higher (e.g. \citealt{HelledGuillot2013,Wahl2017,Helled2018}). When initiating the planetesimals inside the feeding zone of the planets, we reached a total heavy-element mass of $14.8$ and $25.1\, \textrm{M}_{\oplus}$ for Jupiter in scheme 1 and 2, respectively. The same numbers for Saturn were $19.0$ and $24.4\, \textrm{M}_{\oplus}$; however, note that most of the solid mass comes from pebble accretion (the contribution from planetesimal accretion to these numbers was very small, maximally 12\%). Considering scheme 2, we nearly reached the lower end of the predicted heavy-element mass; however, note that this is under the assumption that all planetesimals formed within the feeding zone of the planet, and that the entire pebble flux was converted into planetesimals. When the planetesimals were initiated beyond the feeding zone of the planets, the amount of planetesimal accretion became negligible and the total heavy element mass was almost identical to the pebble isolation mass. 

Based on the above discussion, it is difficult to explain the large heavy-element contents of Jupiter and Saturn with planetesimal accretion during the gas accretion phase. In the following we ask ourselves what are potential ways to increase the accreted planetesimal mass compared to our model? 
\begin{itemize}

\item If the protoplanetary disk mass was increased there would be more solids available, assuming the same solid-to-gas ratio. However, migration and gas accretion are accelerated in massive disks, meaning that the planets need to form later during disk evolution. As a consequence, planetesimal formation at the gap edges initiates later, and the total mass of planetesimals formed at gap edges would not necessarily be larger than in the case with a lower disk mass.  

\item Increasing the solid-to-gas ratio above the standard 1\% would result in more available planetesimal mass. However, in our models we end up with more than $10\, \textrm{M}_{\oplus}$ of planetesimals in a disk beyond Saturn (see Sect. \ref{subsec: removed planetesimals}). When the smallest planetesimal sizes are considered, we also inject $3-9\, \textrm{M}_{\oplus}$ of planetesimals into the inner Solar System (see Fig. \ref{fig: massInt4au}, note that this is not the case for $100\, \textrm{km}$-sized planetesimals). If the formed planetesimal mass is increased by e.g.\ a factor of 10, this means that we would en up with several hundred $\textrm{M}_{\oplus}$ of planetesimals beyond Saturn, which might not be consistent with constraints on the masses of the Solar System's scattered disc and the Oort cloud \citep{Brasser2008}. We nevertheless caution that inferring the original planetesimal mass from the current Oort cloud population is challenging and requires a number of assumptions to be made about the size distribution of the small comets that enter the inner Solar System (see \citealt{Brasser2008} for a discussion).

\item Planetesimals formed at planetary gap edges represent a later generation of planetesimals, which did not contribute to the formation of the giant planet cores. Some of the planetesimals which must have formed earlier during disk evolution, as well as planetesimals forming at other locations in the disk, could also exist in the system. Our results show that planetesimal accretion is very inefficient when the planetesimals form beyond the feeding zone of the planet; however, in \citet{Shibata2020}, \citet{Shibata2021} and \citet{Shibata2022} they show that accretion can be significant if the planets are migrating far and shepherding the planetesimals in front of them. Taking into account the formation of planetesimals via other mechanisms interior of the planets might thus result in more accretion, although additional effects such as ablation needs to be considered when icy planetesimals enter the inner Solar System \citep{Eriksson2021}.

\item The accretion of solids onto the planetary envelope could have an effect on the timescale for gas accretion, which would result in that the planetary growth-tracks changes compared to our model. If the rate of solid accretion during the attached phase is high, then runaway gas accretion can be delayed, resulting in a longer enhanced envelope phase and more planetesimal accretion \citep{Alibert2018,VallettaHelled2020}. The dependence of gas accretion on enrichment is discussed in Sect. \ref{subsubsec: dependence of gas-accretin on enrichment}.

\item If the planetary cores formed much earlier than in our model, regardless of whether it occurred via pebble accretion or via some other scenario (e.g. \citealt{KobayashiTanaka2021}), there would be more solids remaining in the disk and thus the total mass of planetesimals forming at the gap edges would be larger. Our models for gas accretion and migration do not manage to produce Jupiter and Saturn in this scenario, as the planets would become far too massive. The accreted planetesimal mass would likely increase if the cores formed earlier.

\end{itemize}

\subsection{Implications for exoplanets}
The estimated heavy-element mass for hot Jupiters ranges from ${\sim} 10-100\, \textrm{M}_{\oplus}$, with many planets containing more than $50\, \textrm{M}_{\oplus}$ (\citealt{Guillot2006,MillerFortney2011,Thorngren2016}). Although our work focuses on the formation of the cold Solar System giants, it is clear from our results that such large heavy element masses are difficult to explain by the accretion of planetesimals. Despite the fact that all planetesimals were initiated inside the feeding zone of the planet, and we used a 100\% pebble-to-planetesimal conversion efficiency at the gap edges, the maximum accretion efficiency obtained in our simulations was $< 10\%$. In order to accrete ${\sim} 50\, \textrm{M}_{\oplus}$ of planetesimals, there would thus need to be more than $500\, \textrm{M}_{\oplus}$ of planetesimals forming at the gap edges.
 
Based on the discussion in Sect. \ref{subsec: implications to SS formation}, the most promising way of increasing the planetesimal accretion efficiency is to have a prolonged attached phase. One way to achieve this is if the bombardment of planetesimals onto the planetary envelope is high enough to itself delay runaway gas accretion \citep{Alibert2018,VenturiniHelled2020}. Having a longer attached phase also implies a longer migration distance, such that the planet would start accreting gas further out in the disk. This is in line with the results by \citet{Shibata2020}, who found that a Jupiter mass planet can accrete enough planetesimals to explain the observed metallicities, provided that the planet starts migrating at a few tens of au in a massive planetesimal disk (${\sim} 100\, \textrm{M}_{\oplus}$). Efficient planetesimal accretion might thus not be impossible, but it requires extreme conditions. In other cases, alternative processes are required to explain the high metal content of extrasolar giants (see Section \ref{subsec: alt models for giant metallicities}).

\subsection{Alternative models for envelope enrichment}\label{subsec: alt models for giant metallicities}
There are multiple alternative processes that can lead to envelope enrichment, such as accretion of enriched gas, erosion of the initial core or giant impacts. The accretion of enriched gas is a natural outcome of planet formation models, where drifting pebbles sublimate at snow lines and subsequently enrich the gas that is closer to the star (\citealt{Booth2017,SchneiderBitsch2021a,SchneiderBitsch2021b}). Giant planets forming within the snow lines accrete this metal-rich gas and automatically obtain enriched envelopes. In \citet{SchneiderBitsch2021a,SchneiderBitsch2021b} they showed that this process can result in heavy element contents that match the ones predicted by \citet{Thorngren2016} for hot Jupiter systems, provided that the solid-to-gas ratio in the disk is ${\sim} 2\%$. 

Erosion of the initially compact core and subsequent mixing of the heavy elements within the envelope would also result in an enhanced envelope metallicity (e.g. \citealt{Madhusudhan2017}). However, the efficiency of this mechanism is uncertain \citep{Guillot2004} and depends both on material properties and the mixing efficiency within the envelope (\citealt{WilsonMilitzer2012,SoubiranMilitzer2016}). Furthermore, in order to match the estimated heavy element content of exogiants the initial core mass would have to be very large. On the positive side, if feasible, the resulting interior profile of the planet could be one with an extended diluted core, which is typically favored by internal structure models (e.g. \citealt{Wahl2017}).

Finally, studies by \citet{Li2010} and \citet{LiuHori2019} show that energetic head-on collisions between proto-Jupiter and large planetary embryos (several $\textrm{M}_{\oplus}$) could result in shattering/erosion of Jupiter's core. The subsequent mixing of heavy elements into the proto-envelope lead to an enhanced envelope metallicity, and could produce a diluted core profile. The major question mark for this mechanism is whether the frequency of giant impacts is high enough.

\subsection{Shortcomings of the model}
\subsubsection{The effect of gas accretion on disk evolution}
In our model for gas accretion we did not consider the effect of gas accretion on disk evolution, other than the opening of a gap. This is a common simplification in planet formation studies; however, in reality the gas which is accreted onto the planet should be removed from the disk accretion rate onto the star. Consequently, if the disk contains multiple planets that are accreting gas simultaneously, then the amount of gas which is available to the inner planet will depend on how much gas that has been accreted by the outer one. In our model gas accretion onto Jupiter and Saturn are treated independently, which could in practice result in that the sum of the gas accretion rates onto both planets is larger than the disk accretion rate.

Taking the above effects into consideration would result in less gas drag on the planetesimals that are scattered interior of the planetary orbits. This could for example effect the population of circularized planetesimals interior of Jupiter that can be seen in panel 2 and 4 of Fig. \ref{fig: eVSa}. In the case of Jupiter's formation, when Saturn reaches runaway gas accretion the mass available for Jupiter to accrete would decrease compared to the current model. Therefore, Jupiter would have to reach the pebble isolation mass earlier during disk evolution. This would not result in a prolonged enhanced envelope phase (gas accretion is independent of the disk mass during the attached phase); however, given that the pebble flux is larger early on during disk evolution, the accreted planetesimal mass would likely increase a bit.

\subsubsection{The dependence of gas accretion on enrichment}\label{subsubsec: dependence of gas-accretin on enrichment}
The accretion of solids onto the planetary envelope affects the efficiency of gas accretion in mainly two ways: (1) the envelope obtains thermal support from the dissipation of kinetic energy from infalling solids, which counteracts the gravity of the core and thus slows down envelope contraction \citep{Alibert2018}; (2) the enriched envelope obtains a higher molecular weight, which has been shown to result in faster gas accretion and shorter formation timescales (e.g. \citealt{Stevenson1982,Venturini2016,VallettaHelled2020}). If the first effect dominates, the onset of runaway gas accretion is delayed, resulting in a longer migration distance as well as a longer enhanced envelope phase. This would likely lead to a higher accretion efficiency of planetesimals. In \citet{Alibert2018} they find that a constant accretion rate of at least $10^{-6}\, \textrm{M}_{\odot}\, \textrm{yr}^{-1}$ is required in order to stall runaway gas accretion for $2\, \textrm{Myr}$. In principle, the solutions we find with a high planetesimal accretion rate could delay runaway gas accretion and be more consistent with these results. 

If the second effect dominates, gas accretion occurs on a shorter timescale, meaning that the planets in our model would need to reach the pebble isolation mass later during disk evolution. This would likely result in a smaller accreted planetesimal mass. The question of how gas accretion is affected by solid enrichment is an important problem, which should be solved in a more self-consistent manner, with detailed calculations and a proper handling of thermal dynamics

\section{Conclusion}\label{setc: conclusion}
In this work, we study collisions between gap-opening planets and planetesimals forming at their gap edges, with the aim to determine whether the delivered mass is high enough to explain the large heavy element contents of giant planets. To this end, we used a suite of {\it N}-body simulations. We considered the formation of Jupiter and Saturn, taking into account the enhanced collision cross-section caused by their extended envelopes. Two formation pathways were examined, where one leads to large scale planet migration, and the other to migration over a few au only. We further varied the formation location of the planetesimals relative to the planets, and the planetesimal sizes. We find that:

\begin{itemize}
    \item The close proximity to the gap-opening planets causes the planetesimals to leave their birth location soon after formation. Most planetesimals that do not suffer ejection or accretion eventually become members of Saturn's scattered disk. In the case of small planetesimal radii, there is also a population of circularized planetesimals in the innermost disk region.
    \item Planetesimal accretion during the gas accretion phase of giant planet formation is a very inefficient process. The maximum obtained accretion efficiency onto Jupiter or Saturn is less than $10\%$, corresponding to a mass of ${\sim} 3\, \textrm{M}_{\oplus}$ and ${\sim} 2\, \textrm{M}_{\oplus}$, respectively. Since these numbers are obtained assuming that all planetesimals form within the feeding zone of the planets, and that all pebbles reaching the gap-edges are turned into planetesimals, they represent an upper limit on the accreted planetesimal mass. 
    \item When planetesimal formation occurs beyond the feeding zone of the planets, accretion becomes negligible. This is a good indication that planetesimal accretion during the gas-accretion phase will be inefficient also if the planetesimals form via other processes and in other regions of the disk. 
    \item Decreasing the planetesimal radii does not lead to more efficient accretion. In the literature it is often mentioned that having smaller planetesimals leads to more accretion (e.g. \citealt{Alibert2018}), but our results demonstrate that this is not the case for planetesimal accretion during the gas accretion stage. 
    \item The accretion efficiency is higher when we consider the formation pathway with a long migration distance. This is in line with the results by \citet{Shibata2020}, who found that efficient accretion can occur if a massive planet starts migrating at a few tens of au in a massive planetesimal disk (${\sim} 100\, \textrm{M}_{\oplus}$).
\end{itemize}

Based on our results we conclude that it is difficult to explain the large heavy element contents of giant planets with planetesimal accretion during the gas accretion phase, provided they do not migrate very far in a very massive planetesimal disk \citep{Shibata2020}. Hence, alternative processes of envelope enrichment most likely are required in order to explain the high heavy element content inferred for Jupiter and Saturn, as well as that of transiting planets. The accretion of vapor deposited by drifting pebbles is one such promising mechanism \citep{SchneiderBitsch2021a,SchneiderBitsch2021b}.

\begin{acknowledgements}
L.E. and A.J. are supported by the Swedish Research Council (Project Grant 2018-04867). T.R. and A.J. are supported by the Knut and Alice Wallenberg Foundation (Wallenberg Academy Fellow Grant 2017.0287). A.J. further thanks the European Research Council (ERC Consolidator Grant 724 687-PLANETESYS), the G\"{o}ran Gustafsson Foundation for Research in Natural Sciences and Medicine, and the Wallenberg Foundation (Wallenberg Scholar KAW 2019.0442) for research support. R.H. and C.V.~ acknowledge support from the Swiss National Science Foundation (SNSF) under grant 200020\_188460. Simulations in this paper made use of the REBOUND code which is freely available at http://github.com/hannorein/rebound. The computations were performed on resources funded by the Royal Physiographic Society of Lund.
\end{acknowledgements}

% References 
\bibliographystyle{aa} % style aa.bst
\bibliography{ref} % your references Yourfile.bib

\appendix

\section{Details of the disk model}\label{sec: appendix disk model}
We use the analytic solution for the surface density evolution of an unperturbed thin accretion disk from \citet{LyndenBellPringle1974}
\begin{equation}\label{eq: sigmaUnp}
\Sigma_{\rm unp}(t)=\frac{\dot{M}_g(t)}{3\pi \nu_{\rm out} (r/r_{\rm out})^{\gamma}} \exp \left[-\frac{(r/r_{\rm out})^{(2-\gamma)}}{T_{\rm out}}\right],
\end{equation}
where
\begin{equation}
T_{\rm out}=\frac{t}{t_s}+1,
\end{equation}
and
\begin{equation}
t_s = \frac{1}{3(2-\gamma)^2}\frac{r_{\rm out}^2}{\nu_{\rm out}}.
\end{equation}
In the equations above $\dot{M}_g(t)$ is the disk accretion rate at time $t$, $r$ is the semimajor axis, $r_{\rm out}$ is the scaling radius, $\nu_{\rm out} \equiv \nu(r_{\rm out})$ where $\nu$ is the kinematic viscosity and $\gamma$ is the radial gradient of $\nu$. The evolution of the disk accretion rate is given by
\begin{equation}
\dot{M}_g(t) = \dot{M}_{g}(t=0) \left[\frac{t}{t_s}+1 \right]^{ -(\frac{5}{2}-\gamma)/(2-\gamma) } - \dot{M}_{\rm pe},
\end{equation}
where $\dot{M}_{\rm pe}$ is the rate at which material is removed by photoevaporation, which we take to be constant in time.

The sound speed in the disk is calculated as
\begin{equation}\label{eq:cs}
c_{\textrm{s}} = \left(\frac{k_{\textrm{B}} T}{\mu m_{\textrm{H}}}\right)^{1/2},
\end{equation}
where $k_{\textrm{B}}$ is the Bolzmann constant, $T$ is the temperature, $\mu$ is the mean molecular weight, and $m_{\textrm{H}}$ is the mass of the hydrogen atom. We use a value of $2.34$ for the mean molecular weight \citep{Hayashi1981}. The midplane temperature of the disk was approximated using a fixed powerlaw structure: 
\begin{equation}\label{eq:Tdisc}
T=150\, \textrm{K} \times (r/{\rm au})^{-3/7}
\end{equation}
\citep{ChiangGoldreich1997}.

We used Gaussian gap profiles to model the planetary gaps, where the Gaussian is described by the equation
\begin{equation}
G(r) = \left(1-\frac{\Sigma_{\textrm{gap}}}{\Sigma_{\rm unp}}\right) \exp \left[-\frac{(r-a)^2}{2H_a^2}\right],
\end{equation}
where $a$ is the semimajor axis of the planetary orbit and $H_a$ is the corresponding gas scale height. The perturbed surface density profile could then be obtained by using the following expression:
\begin{equation}
\Sigma = \frac{\Sigma_{\rm unp}}{1+G_{\textrm{1}}+G_{\textrm{2}}+...},
\end{equation}
where each planet contributes their own Gaussian. Finally, the gas density at some height, $z,$ away from the midplane was obtained by using the equation
\begin{equation}
\rho(z) = \frac{\Sigma}{\sqrt{2\pi}H} \exp \left[\frac{-z^2}{2H^2}\right],
\end{equation}
where we have assumed vertical hydrostatic equilibrium for the gas in the disk.

\section{Gas accretion rate}
\begin{figure}
\centering
\includegraphics[width=\hsize]{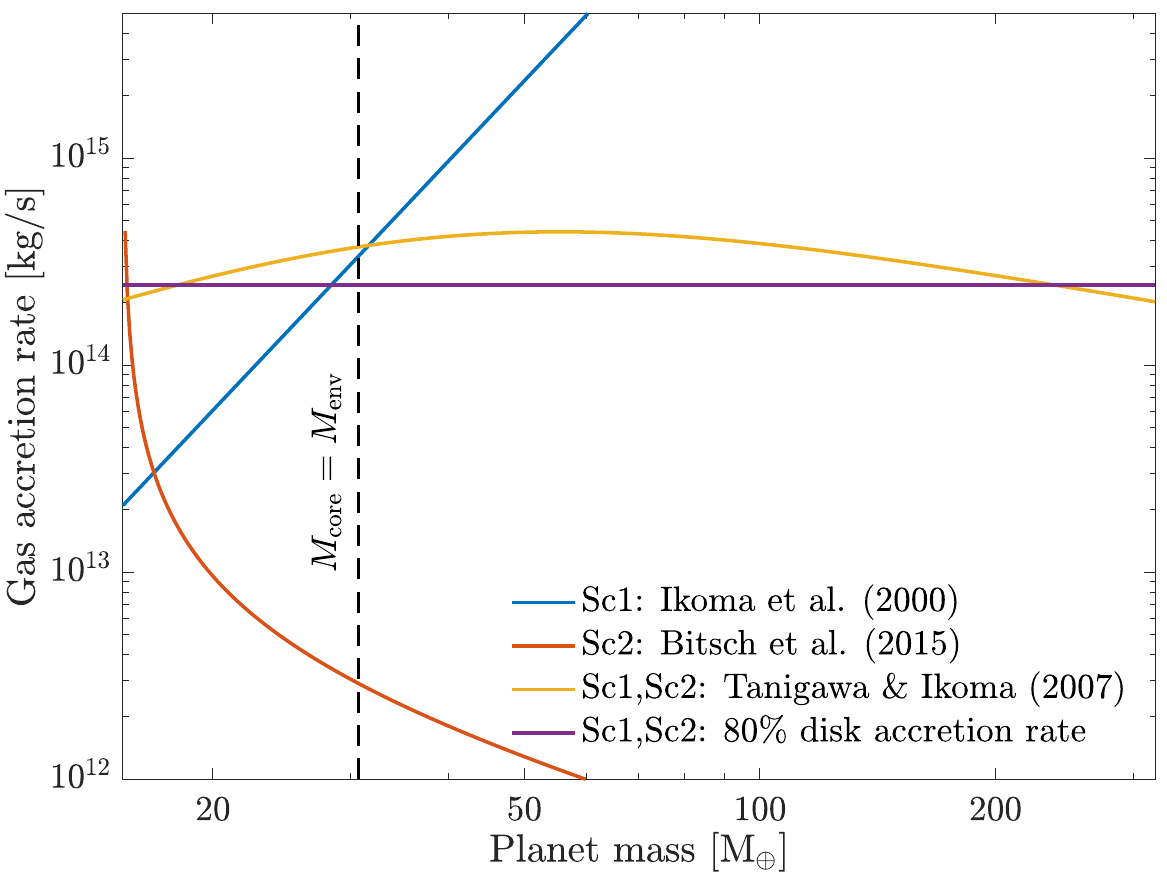}
\caption{Gas accretion rate as a function of planetary mass for a planet located at $10\, \textrm{au}$ in a $2\, \textrm{Myr}$ old protoplanetary disk. The labels Sc1 and Sc2 tells in which gas accretion scheme the current model is being used. }
\label{fig: gasAccretion}
\end{figure}

In Fig. \ref{fig: gasAccretion} we have plotted the dependence on planetary mass for all the different gas accretion rates which are being used in our models. The rates are calculated using a semimajor axis of $10\, \textrm{au}$ and a disk time of $2\, \textrm{Myr}$. In scheme 1 the gas accretion rate during envelop contraction increases with increasing planetary mass, and runaway gas accretion initiates when this rate becomes higher than what the disk can supply. In scheme 2 the gas accretion rate during envelop contraction instead decreases with increasing planetary mass, and runaway gas accretion initiates when the core mass equals the envelope mass. At the chosen time and semimajor axis, gas accretion is limited by disk accretion for most of the runaway phase. 

\section{Additional plots}\label{sec: appendix planetesmal accretion}
\begin{figure*}
\centering
{\includegraphics[width=\columnwidth]{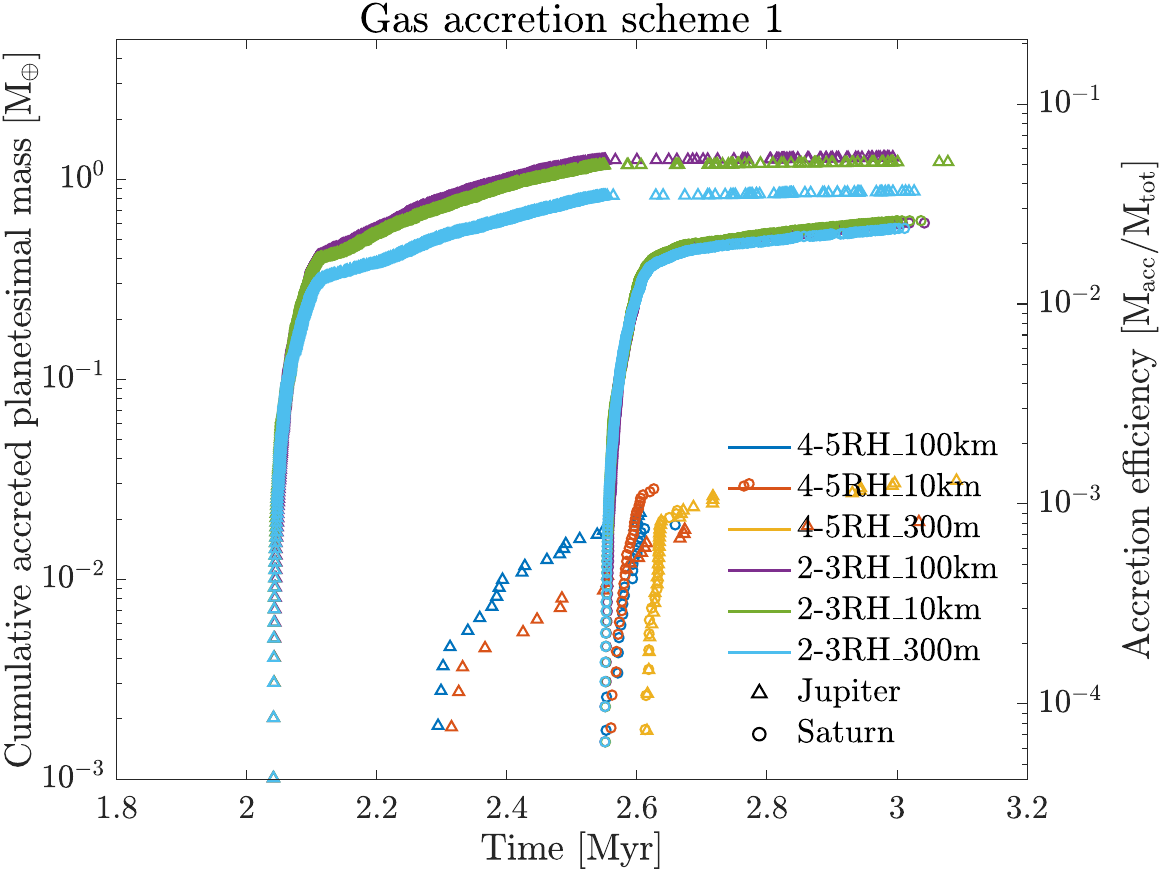}}
{\includegraphics[width=\columnwidth]{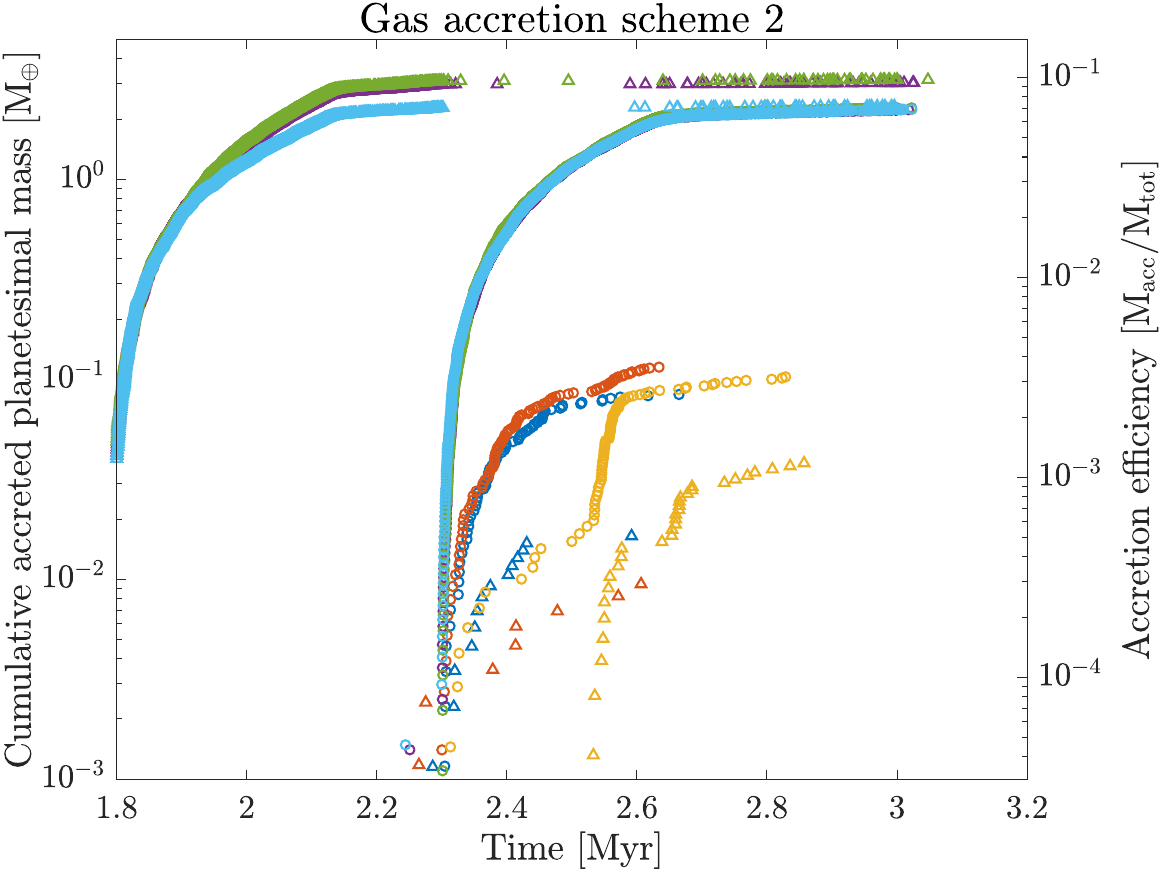}}
\caption{\textit{Left:} Cumulative accreted planetesimal mass as a function of time for all simulations in scheme 1. The corresponding accretion efficiency is calculated by dividing with the total planetesimal mass that has formed in the system at the time of disk dissipation. Each scatter point in the plot is one collision event; however, since we performed three simulations per parameter set and show the combined results, the number of collisions in one simulation should be three times smaller than in this plot. The maximum obtained accretion efficiency onto any planet is 5\%, which corresponds to ${\sim} 1\, \textrm{M}_{\oplus}$. The accretion efficiency is much higher for planetesimals formed inside the feeding zone, and does not change significantly with planetesimal size. \textit{Right:} Same as the left plot but for gas accretion scheme 2. The maximum accretion efficiency obtained for any planet is $10\%$, which corresponds to a mass of ${\sim} 3\, \textrm{M}_{\oplus}$. }
\label{fig: planAcc}
\end{figure*}

In Fig. \ref{fig: planAcc} we show the amount of planetesimal mass that has been accreted onto Jupiter and Saturn as a function of time for all simulations, along with the corresponding accretion efficiency. This is the same data that is presented in Fig. \ref{fig: colDiag}. The "knee" on the accretion curves for planetesimals formed within the feeding zone coincides with the location where $M_{\rm env}=M_{\rm core}$, which is when runaway gas accretion initiates and the approximation for the capture radius becomes significantly smaller (see right panel of Fig. \ref{fig: growthTrack} for a plot of the capture radius versus time). Up until this point a fraction of the planetesimals are captured immediately after formation due to the enhanced envelope. This trend is not seen in the accretion curves for planetesimals formed beyond the feeding zone, since they are initiated too far away from the planet to be affected by the enhanced envelope.

\begin{figure*}
\centering
{\includegraphics[width=\columnwidth]{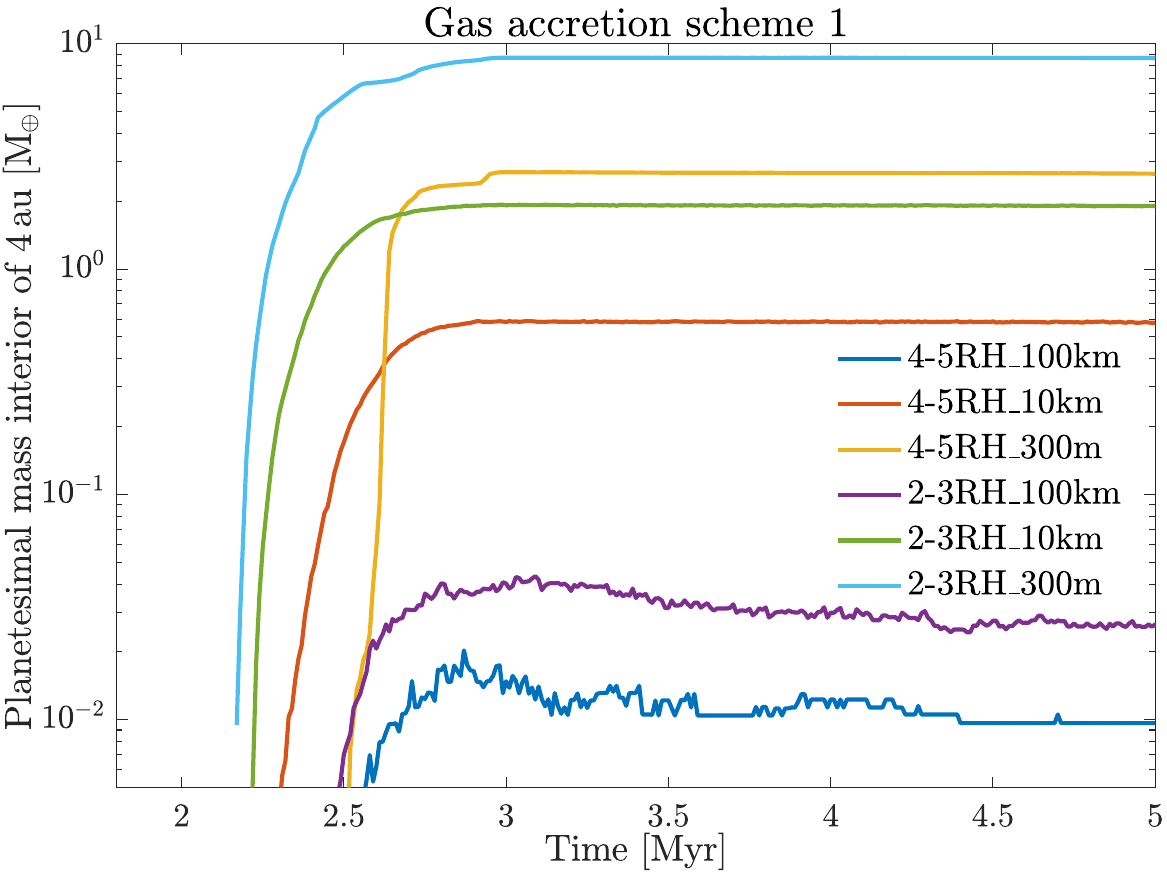}}
{\includegraphics[width=\columnwidth]{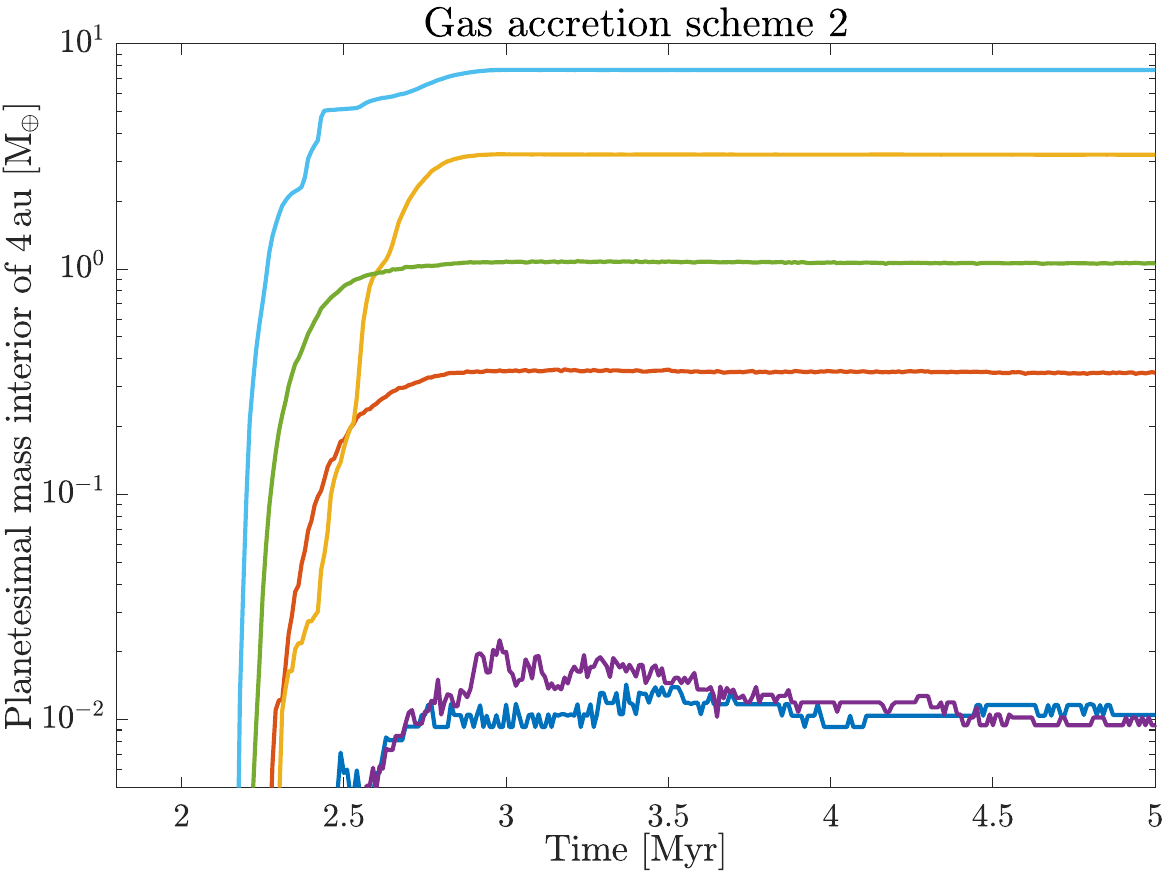}}
\caption{\textit{Left:} The total planetesimal mass residing in the inner Solar System (planetesimals with aphelion less than $4\, \textrm{au}$) plotted as a function of time for all simulations in scheme 1. The chance of making it into the inner Solar System increases with decreasing planetesimal size and decreasing formation distance relative to the planet. \textit{Right:} Same as the left plot but for gas accretion scheme 2.  }
\label{fig: massInt4au}
\end{figure*}

In Fig. \ref{fig: massInt4au} the total planetesimal mass residing in the inner Solar System is shown as a function of time (we consider a planetesimal to be in the inner Solar System if its aphelion is less than $4\, \textrm{au}$). In the case of $100\, \textrm{km}$-sized planetesimals, the amount of mass injection into the inner Solar System is negligible. Considering $10\, \textrm{km}$-sized planetesimals, we inject between $0.3-2\, \textrm{M}_{\oplus}$ of planetesimals into the inner Solar System, which remains also after the dispersal of the gas disk. When using planetesimals of $300\, \textrm{m}$ in size, we scatter and subsequently trap $3-9\, \textrm{M}_{\oplus}$ of planetesimals in the inner Solar System. We find that planetesimals forming inside the feeding zone of the planet has a higher chance of making it into the inner Solar System, which is because of the more efficient planetary scattering. If the amount of planetesimal formation at the gap edges were to drastically increase, the amount of mass entering the inner Solar System would do so as well.

\begin{figure}
\centering
{\includegraphics[width=\columnwidth]{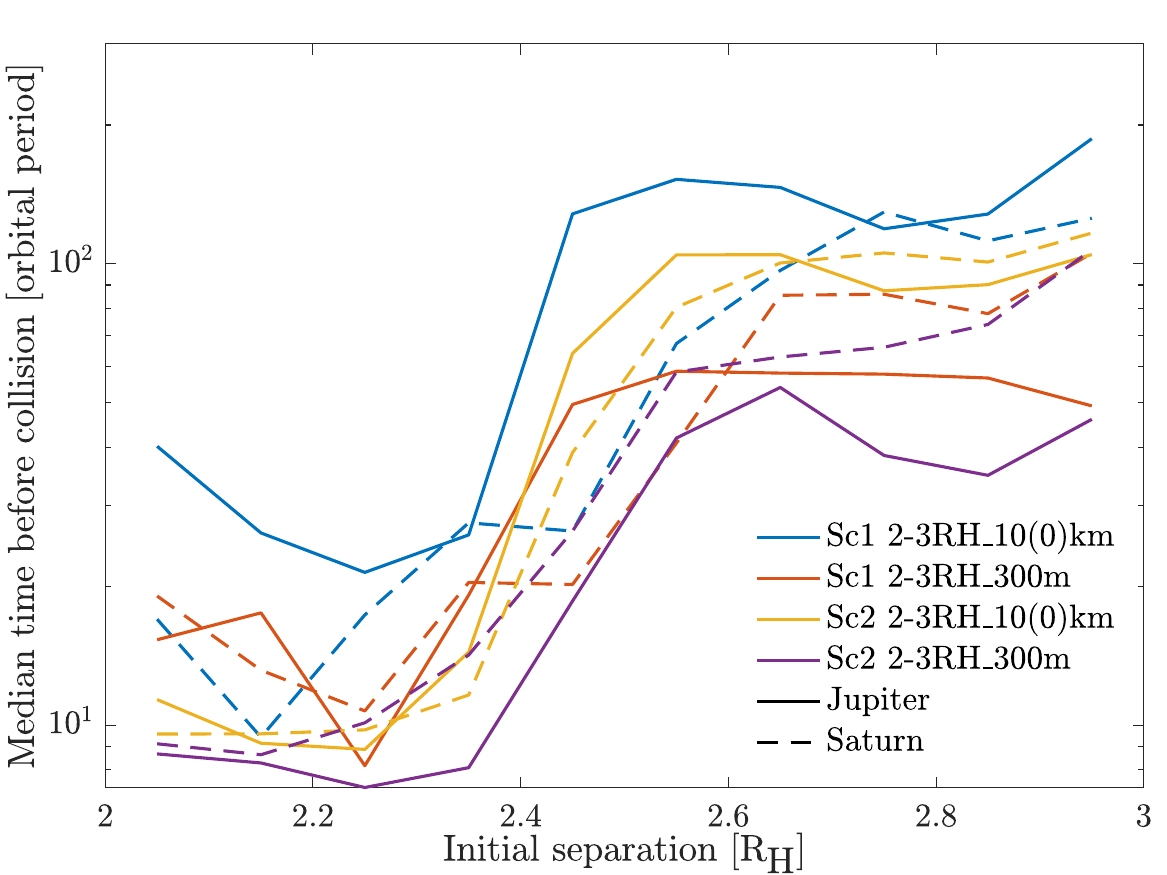}}
\caption{Median time before collision versus the initial separation for all simulations with planetesimals initiated inside the feeding zone of the planets. The time is given in units of the orbital period of the planets at the time the planetesimals are formed. Planetesimals initiated between $2-2.4\, R_H$ typically collide within 20 orbital periods.}
\label{fig: collTime_formLoc}
\end{figure}

In Fig. \ref{fig: collTime_formLoc} the median time before collision is presented as a function of the initial separation relative to the planet. Planetesimals that are initiated between $2-2.4\, R_H$ typically collide within 20 orbital periods. This suggests that a significant fraction of the accretion efficiency comes from immediate accretion (a 10 orbital period lifetime is roughly 2-3 close encounters). Whether or not it is realistic for planetesimals to form in regions with immediate accretion we do not know, and has to our knowledge not been studied.

\end{document}